\documentclass[traditabstract]{aa}

\usepackage{graphicx}
\usepackage{txfonts}
\usepackage{bm}

\usepackage{natbib}
\bibpunct{(}{)}{;}{a}{}{,}

\newcommand{\beq}{\begin{equation}}
\newcommand{\eeq}{\end{equation}}
\newcommand{\bea}{\begin{eqnarray}}
\newcommand{\eea}{\end{eqnarray}}
\newcommand{\req}[1]{Eq.~\ref{#1}}
\newcommand{\ael}{a_\mathrm{e}}
\newcommand{\aion}{a_\mathrm{i}}
\newcommand{\alphaf}{\alpha_\mathrm{f}}
\newcommand{\am}{a_\mathrm{m}}
\newcommand{\chie}{\chi_\mathrm{e}}
\newcommand{\dd}{\mathrm{d}} 

\newcommand{\Gamme}{\Gamma_\mathrm{e}}
\newcommand{\gammam}{\gamma_\mathrm{m}}
\newcommand{\Gammam}{\Gamma_\mathrm{m}}
\newcommand{\gcc}{\mbox{g~cm$^{-3}$}}
\newcommand{\gfact}{g_\mathrm{i}}
\newcommand{\gr}{\gamma_\mathrm{r}}
\newcommand{\EF}{\epsilon_\mathrm{F}}
\newcommand{\lambde}{\lambda_\mathrm{e}}
\newcommand{\lambdi}{\lambda_\mathrm{i}}
\newcommand{\mel}{m_\mathrm{e}}
\newcommand{\mion}{m_\mathrm{i}}
\newcommand{\Mspin}{M_\mathrm{spin}}
\newcommand{\nel}{n_\mathrm{e}}
\newcommand{\mue}{\mu_\mathrm{e}}
\newcommand{\Nel}{N_\mathrm{e}}
\newcommand{\nion}{n_\mathrm{i}}
\newcommand{\Nion}{N_\mathrm{i}}
\newcommand{\kB}{k_\mathrm{B}}

\newcommand{\omp}{\omega_\mathrm{p}}
\newcommand{\omc}{\omega_\mathrm{c}}
\newcommand{\omci}{\omega_\mathrm{ci}}
\newcommand{\pF}{p_\mathrm{F}}
\newcommand{\Prel}{P_\mathrm{r}}
\newcommand{\rhos}{\rho_\mathrm{s}}
\newcommand{\rs}{r_\mathrm{s}}
\newcommand{\RS}{R_\mathrm{S}}
\newcommand{\TF}{T_\mathrm{F}}

\newcommand{\Tm}{T_\mathrm{m}}
\newcommand{\Tp}{T_\mathrm{p}}
\newcommand{\Tr}{T_\mathrm{r}}
\newcommand{\Uq}{U_\mathrm{q}}
\newcommand{\xr}{x_\mathrm{r}}
\newcommand{\zeti}{\zeta_\mathrm{i}}

\begin{document}


\title{Equation of state for magnetized Coulomb plasmas}

\author{A. Y. Potekhin
  \inst{1,2,3}
  \thanks{\email{palex@astro.ioffe.ru}}
\and
G. Chabrier
  \inst{1,4}
  \thanks{\email{chabrier@ens-lyon.fr}}
  }

\institute{CRAL (UMR CNRS No. 5574), 
Ecole Normale Sup\'{e}rieure de Lyon,
69364 Lyon Cedex 07, France
\and
Ioffe Physical-Technical Institute,
Politekhnicheskaya 26, 194021 St.~Petersburg, Russia
\and
Isaac Newton Institute of Chile, 
         St.~Petersburg Branch, Russia
\and
School of Physics, University of Exeter, Exeter, UK EX4 4QL}

\date{Received 23 July 2012 / Accepted 5 December 2012}

\abstract{We have developed an analytical
equation of state (EOS) for magnetized fully-ionized plasmas
that cover
a wide range of temperatures and densities, from low-density classical
plasmas to relativistic, quantum plasma conditions. This EOS directly applies to
calculations of structure and evolution of strongly
magnetized white dwarfs and neutron stars.
We review available analytical and
numerical results for thermodynamic functions of the
nonmagnetized and
magnetized Coulomb
gases, liquids, and solids. 
We propose a new analytical expression for the free
energy of solid Coulomb
mixtures. Based on recent numerical results, we
have constructed analytical approximations for the
thermodynamic functions of harmonic Coulomb crystals in
quantizing magnetic fields.
The analytical description ensures
a consistent evaluation of all astrophysically important
thermodynamic functions based on the first, second, and
mixed derivatives of the free energy. Our numerical code for
calculation of thermodynamic functions based on these
approximations has been made publicly available.\thanks{The
Fortran subroutines are available at the CDS via
anonymous ftp to cdsarc.u-strasbg.fr (130.79.128.5)
or via http://cdsweb.u-strasbg.fr/cgi-bin/qcat?J/A+A/.}
Using this code, we calculate and discuss the effects
of electron screening and magnetic quantization on the position of
the melting point in a range of densities and magnetic
fields relevant to white dwarfs and outer envelopes of
neutron stars. We consider also the thermal and mechanical structure of
a magnetar envelope and argue that it can have a frozen
surface which covers the liquid ocean above the solid
crust.}

\keywords{dense matter -- equations of state -- magnetic
fields -- stars: neutron -- white dwarfs}

\maketitle

\section{Introduction} 
\label{sect:intro}

Coulomb plasmas, i.e., the fully-ionized plasmas whose
thermodynamics is strongly affected by electrostatic
interactions, are encountered in many physical and astrophysical
situations \citep[e.g.,][]{Fortov}. Full ionization is
reached either at high temperatures $T$ and low densities
$\rho$ (thermal ionization) or at high densities $\rho$
(pressure ionization). The latter case is typical of the interior conditions
of low-mass stars, brown dwarfs, or giant planets
\citep{ChaBar} as well as
the interior and envelope conditions of
white dwarfs and neutron stars. Coulomb interactions
are crucial for the equation of state (EOS) under such conditions.
In the interior or the envelope of compact objects such as white dwarfs and neutron stars, the
electrons can be weakly or strongly degenerate, the plasma
can be in the liquid or solid state, the electrons can have
various degrees of degeneracy and relativism, and the
quantum effects on ion motion can be substantial. 
Therefore, a wide-range EOS is needed for calculations of
the structure and evolution of such stars.

In a previous work (\citealp{PC00,PC10}, hereafter 
Paper~I
and Paper~II, respectively) we proposed a set of analytical
expressions for the calculations of the EOSs of the Coulomb
plasmas without magnetic fields and presented a code for
thermodynamic functions based on the first, second, and
mixed derivatives of the analytical Helmholtz free energy
$F$ with respect to density $\rho$ and temperature $T$. This
code has been employed in astrophysical modeling and adapted
for the use in the Modules for Experiments in Stellar
Astrophysics (\textsc{MESA}) \citep{MESA}.

The Bohr -- van Leeuwen theorem states that an EOS of
charged pointlike \emph{classical} particles is not affected
by a magnetic field \citep{vanLeeuwen}. However, a magnetic field
can affect thermodynamic functions through intrinsic
magnetic moments of particles and by the quantization of the
motion of charged
particles in Landau orbitals \citep{Landau30,LaLi-QM}.
These effects can be important, for example, in magnetic
white dwarfs whose magnetic fields $B$ can reach 
$10^7-10^9$~G (e.g., \citealp{WickramaFerrario00}, and
references therein) and neutron stars with typical
$B\sim10^{8}-10^{14}$~G (e.g., \citealp{NSB1}, and
references therein).

In this paper, we systematically consider analytical
expressions for thermodynamic functions of  magnetized
Coulomb plasmas, discuss their validity range, and introduce
some practical modifications.  We take account of analytical
and numerical results, currently available for various
contributions to the Helmholtz free energy in quantizing
magnetic fields. Taking advantage of recently published
numerical results \citep{Baiko09}, we construct an
analytical description of the thermodynamic functions of
harmonic Coulomb crystals in quantizing magnetic fields.

In Sect.~\ref{sect:plasmpar} we give definitions and simple
estimates for the plasma parameters that determine different
thermodynamic regimes. In Sect.~\ref{sect:nonmag} we 
outline the EOS of a nonmagnetized Coulomb plasma as
the reference case. In Sect.~\ref{sect:mag-id} we consider 
the Boltzmann and Fermi gases in quantizing magnetic fields,
present a general analytical description of their EOS, and
simplify them for several limiting cases. In
Sect.~\ref{sect:mag-other} we review nonideal contributions
to the EOS of a Coulomb liquid in a strong magnetic field.
In Sect.~\ref{sect:mag-sol} we derive an analytical
approximation for the EOS of a strongly magnetized Coulomb
crystal. In Sect.~\ref{sect:examples} we present and discuss
examples of thermodynamic functions for conditions typical
of white dwarfs and neutron-star envelopes. The summary is
given in Sect.~\ref{sect:concl}.
In the Appendices we give the explicit expressions for
the thermodynamic functions used in
Sect.~\ref{sect:nonmag}.

\section{Basic definitions}
\label{sect:plasmpar}

\subsection{General parameters}
\label{sect:gen}

Let $\nel$ and $\nion$ be the electron and ion number
densities, $A$ and $Z$ the ion mass and charge numbers,
respectively. In this paper we consider the neutral plasmas
(therefore $\nel = Z \nion$) that contain
a single type of ion and include neither positrons (they
can be described using the same analytical functions as for
the electrons; see, e.g., \citealp{Blin,TimmesArnett}), nor
free neutrons (see \citealp{NSB1} for a review).

The state of a free-electron gas is determined by the
electron number density $\nel$ and temperature $T$. Instead
of $\nel$ it is convenient to introduce the dimensionless
density parameter $\rs=\ael/a_0$, where $a_0$ is
the Bohr radius and $\ael=(\frac43\pi \nel)^{-1/3}$. The
parameter $\rs$ can be quickly evaluated from the
relations
$
   \rs = 1.1723 \,n_{24}^{-1/3}=(\rho_0/\rho)^{1/3},
$
where $n_{24} \equiv \nel/10^{24}{\rm~cm}^{-3}$
and $\rho_0=2.6752\,(A/ Z)$ g~cm$^{-3}$.
The analogous density parameter for the
ions is 
$\RS=\aion \mion (Ze)^2/\hbar^2 = 1822.89AZ^{7/3}\,
\rs$,
where $\mion$ is the ion mass
and $\aion\equiv(\frac43\pi \nion)^{-1/3}$ is the ion sphere radius.

At stellar densities it is convenient to use, instead of $\rs$,
the nonmagnetic relativity parameter
\begin{equation}
    \xr  = \pF / \mel c = 1.00884 \,
       \left( \rho_6  Z / A 
       \right)^{1/3}\!\! = 0.014005\,\rs^{-1},
\label{x_r}
\end{equation}
where $ \pF = \hbar \,(3 \pi^2 \nel)^{1/3}$
is the electron Fermi momentum in the absence of a magnetic
field, and $\rho_6\equiv\rho/10^6$ g~cm$^{-3}$.
The Fermi energy (without the rest energy) for the electron gas is
$\EF = c\,\sqrt{(\mel c)^2 + (\pF)^2}-\mel c^2 ,
$
and the Fermi temperature $\TF \equiv \EF/\kB
= \Tr  \, ( \gr - 1) ,$
where 
$\Tr \equiv {\mel c^2 / \kB } = 5.93\times
10^9~\mathrm{K}$,
$\gr \equiv \sqrt{1+ \xr^2}$,
and $\kB$ is the Boltzmann constant.
A useful measure of electron degeneracy is $\theta=T/\TF$.
In the nonrelativistic limit ($\xr\ll1$),
$\TF\approx 1.163\times10^6\, \rs^{-2}$~K,
and
\beq
   \theta = 0.543\,\rs/\Gamme,
\label{theta}
\eeq
where
\beq
    \Gamme \equiv \frac{ e^2 }{ \ael \kB T}
   \approx \frac{22.747}{ T_6}
       \left(\rho_6 \frac{  Z  }{ A}
       \right)^{1/3}.
\eeq
In the opposite ultrarelativistic limit ($\xr\gg1$),
$\theta\approx(263\,\Gamme)^{-1}$.
The strength of the Coulomb interaction of nonrelativistic ions
is characterized by the Coulomb coupling parameter
\begin{equation}
   \Gamma =
        \frac{(Z e)^2}{\aion  \kB T} = \Gamme Z^{5/3},
\end{equation}
where $T_6\equiv T/10^6$~K.

Thermal de Broglie wavelengths of free ions and electrons are
usually defined as
\beq
   \lambdi = \left(\frac{2\pi\hbar^2}{ \mion \kB T}\right)^{1/2},
\quad
   \lambde = \left(\frac{2\pi\hbar^2}{ \mel \kB
          T}\right)^{1/2}\,,
\label{lambda-th}
\end{equation}
although in some publications these definitions differ by a
numerical factor.
The quantum effects on ion motion are important
either at $\lambdi\gtrsim \aion$ or at 
$T \ll \Tp$, where $\Tp \equiv \hbar \omp/ \kB$
is the ion plasma temperature and
$
    \omp = \left(  {4 \pi e^2 \,\nion}
     Z^2 /\mion \right)^{1/2}
$
is the ion plasma frequency.
Since $\lambdi/\aion=(\Tp/T)\sqrt{2\pi/3\Gamma}$, 
the importance of the quantum effects in strongly coupled
plasmas (i.e., at $\Gamma\gg1$) is determined by parameter
\beq
   \eta \equiv \Tp/T = \Gamma\sqrt{3/\RS}
       \approx 7.832\, (Z/A)\sqrt{\rho_6}/T_6.
\eeq

\subsection{Magnetic-field parameters}
\label{sect:mag-par}

In the nonrelativistic theory \citep{LaLi-QM}, the energy of
an electron in magnetic field $\bm{B}$ equals $n\hbar\omc+\mel
p_z^2/2$, where $p_z$ is the momentum component along
$\bm{B}$, $\omc=eB/\mel c$ is the electron cyclotron
frequency, $n=n_\mathrm{L}+\frac12-\frac12\sigma$
characterizes a Landau level, $\sigma=\pm1$ determines
the spin projection on the field, and $n_\mathrm{L}$ is the
non-negative integer Landau number related to the
quantization of the kinetic motion transverse to the field.
In  the relativistic theory
\citep{JohnsonLippmann,LaLi-QED}, the kinetic energy
$\epsilon$ of an electron
at the Landau level $n$ and its longitudinal momentum $p_z$
are inter-related as
\bea
&&
   \epsilon = \epsilon_n(p_z) = c\,\left(\mel^2 c^2 + 2\hbar\omc
   \mel n+p_z^2\right)^{1/2} - \mel c^2,
\label{magnenergy}
\\&&
   |p_z| = p_n(\epsilon) = \left[
              (\mel c + \epsilon/c)^2 - (\mel c)^2
                      - 2 \mel \hbar \omc n  \right]^{1/2}.
\label{magnmoment}
\eea
The levels $n\geqslant 1$ are double-degenerate with respect
to $\sigma$. Their splitting due to the anomalous magnetic
moment of the electron is $\approx(\mel c^2
\alphaf/2\pi)\,b$ at $b\ll1$ and $\sim(\mel
c^2\alphaf/2\pi)\,[\ln b-1.584]^2$ at $b\gg1$
\citep[see][]{Schwinger88,SuhMathews}, which is
always much smaller than $\hbar\omc$ and is negligible in
the compact stars.

Convenient dimensionless parameters that characterize the
magnetic field in a plasma are the ratios of
the electron cyclotron energy
$\hbar\omc$ to the Hartree unit of energy, to the electron
rest energy, and to $\kB T$:
\beq
   \gammam=\hbar^3 B / \mel^2 c e^3 =
         B/B_0,
\quad
\eeq
where $B_0=2.3505\times10^9\mbox{~G}$,
\beq
   b=\frac{\hbar\omc}{\mel c^2}
       =\alphaf^2\, \gammam =
           \frac{B}{4.414\times10^{13}\mbox{~G}}\,,
\eeq
where $\alphaf=e^2/\hbar c$ is the fine-structure constant,
and
\beq
   \zeta = \hbar\omc/\kB T \approx 134.34 B_{12}/T_6,
\eeq
where
$B_{12}\equiv B/10^{12}$~G.
The magnetic length 
  $\am=(\hbar c/eB)^{1/2}=a_0/\sqrt{\gammam}$  gives a
characteristic transverse scale of the electron wave
function. 

For the ions, the cyclotron energy is $\hbar\omci =
Z\,(\mel/\mion)\, \hbar\omc$, and the parameter analogous to
$\zeta$ is
\beq
   \zeti = \hbar\omci/\kB T \approx 0.0737\,(Z/A)
           B_{12}/T_6.
\label{omci}
\eeq
Another important parameter is the ratio of the ion
cyclotron frequency to the plasma frequency,
\beq
   \beta = \omci/\omp = \zeti/\eta
          \approx 0.0094\,B_{12}/\sqrt{\rho_6}.
\eeq

\subsection{Free energy and thermodynamic functions}

The Helmholtz free energy $F$ of a plasma
can be conveniently written as
\beq
   F = 
   F_\mathrm{id}^\mathrm{(i)} + F_\mathrm{id}^\mathrm{(e)} 
   + F_\mathrm{ee} + F_\mathrm{ii} + F_\mathrm{ie}, 
\label{f-tot}
\eeq
where $F_\mathrm{id}^\mathrm{(i)}$ and
$F_\mathrm{id}^\mathrm{(e)}$  denote the ideal free energy
of the ions and the electrons, and the last three terms
represent an excess free energy arising from  the
electron-electron, ion-ion, and ion-electron interactions,
respectively. In the nonideal plasmas, correlations between
any plasma particles depend on all interactions, therefore
the separation in \req{f-tot} is just a question of
convenience.

An important reference case is the model of one-component
plasma (OCP). In this model, the electrons are replaced by a
rigid (nonpolarizable) background of the uniform charge
distribution. It is convenient to define $F_\mathrm{ii}$ as
the difference between $F$ and $F_\mathrm{id}^\mathrm{(i)}$
in the OCP model. Still stronger simplification is the
ion-sphere model, in which the interaction energy in the OCP
is evaluated as the electrostatic energy of a positive ion
in the negatively charged sphere of radius $\aion$
\citep{Salpeter61}. The electron exchange-correlation term
is defined as  $F_\mathrm{ee}=F-F_\mathrm{id}^\mathrm{(e)}$
in the model of an electron gas without consideration of the
ions, which are replaced by an uniform positive background
to ensure the global charge neutrality. The
ion-electron (electron polarization) contribution
$F_\mathrm{ie}$, then, is the difference between $F$ and the other
terms, when interactions between all types of particles are
taken into account.

The pressure $P$, the internal energy $U$, and the entropy
$S$ of an ensemble of particles in volume $V$ can be
found from the thermodynamic relations
$
   P=-(\partial F/\partial V)_T, \quad
   S= -(\partial F / \partial T )_V,
$
and $U=F+TS$.
The second-order thermodynamic functions are derived by
differentiating these first-order functions. The
decomposition (\req{f-tot}) induces analogous decompositions
of $P$, $U$, $S$, the heat capacity
$C_V=(\partial S/\partial\ln T)_V,$ and
the logarithmic derivatives
$
   \chi_T=(\partial\ln P/\partial\ln T)_V
$ 
and
$
   \chi_\rho=-(\partial\ln P/\partial\ln V)_T.
$
Other second-order functions can be expressed through these
functions by Maxwell relations (e.g., \citealp{LaLi-SP1}).

\section{EOS of nonmagnetized Coulomb plasmas}
\label{sect:nonmag}

\subsection{Ideal part of the free energy}
\label{sect:id}

The free energy of a gas of
$\Nion=\nion V$ nonrelativistic classical ions is
\beq
   F_\mathrm{id}^\mathrm{(i)} =
     \Nion \kB T \left[\ln(\nion\lambdi^3/\Mspin)-1 \right],
\label{id_i}
\end{equation}
where $\Mspin$ is the spin multiplicity. Accordingly,
$U_\mathrm{id}^\mathrm{(i)} =\frac32\, \Nion \kB T$,
$P_\mathrm{id}^\mathrm{(i)} =\nion \kB T$,
$C_{V,\mathrm{id}}^\mathrm{(i)} =\frac32\, \Nion \kB$, and
$\chi_{T,\mathrm{id}}^\mathrm{(i)}=
\chi_{\rho,\mathrm{id}}^\mathrm{(i)}=1$.
In the OCP, \req{id_i} can be written in terms of the
dimensionless plasma parameters (Sect.~\ref{sect:plasmpar}) as
\beq
   \frac{F_\mathrm{id}^\mathrm{(i)}}{\Nion\kB T}
      = 3 \ln \eta - \frac32 \ln \Gamma
        - \frac12 \ln\frac6\pi-\ln \Mspin -1.
\eeq

The free energy of the electron gas is given by
\beq
   F_\mathrm{id}^\mathrm{(e)} =
   \mue \Nel  - P_\mathrm{id}^\mathrm{(e)}\,V,
\label{id_e}
\end{equation}
where $\Nel=\nel V$ is the number of electrons and $\mue$ is
the electron chemical potential without the rest energy
$\mel c^2$. The pressure and the number density are
functions of $\mue$ and $T$:
\bea
   P_\mathrm{id}^\mathrm{(e)} &=&
 \frac{8}{3\sqrt\pi}\,\frac{\kB T }{ \lambde^3}
   \left[ I_{3/2}(\chie,\tau)
   + \frac{\tau}{ 2}I_{5/2}(\chie,\tau) \right],
\label{P_e}
\\
   \nel &=&
           \frac{4}{\sqrt{\pi}\,\lambde^3}
   \left[ I_{1/2}(\chie,\tau)
   + \tau I_{3/2}(\chie,\tau) \right],
\label{n_e}
\eea
where 
$\chie\equiv\mue/\kB T$, $\tau\equiv T/\Tr$, and
\beq
   I_\nu(\chie,\tau) \equiv \int_0^\infty
  \frac{ x^\nu\,(1+\tau x/2)^{1/2}
    }{ \exp(x-\chie)+1 }\,{\dd}x
\label{I_nu}
\eeq
is the Fermi-Dirac integral. The internal energy is
\beq
   U_\mathrm{id}^\mathrm{(e)} =
             \frac{4}{\sqrt\pi}\,\frac{\kB T V}{ \lambde^3}
   \left[ I_{3/2}(\chie,\tau) +
         \tau\,I_{5/2}(\chie,\tau) \right].
\label{U_e}
\eeq
Since we use $V$ and $T$ as independent variables,
we need to find $\mue(V,T)$. This can be done either
by inverting \req{n_e} numerically, or from the analytical
approximation given in \citet{CP98}. Then the second-order
thermodynamic functions are obtained using  relations of the
type
\bea&&
   \left(\frac{\partial f(\chie,T)}{\partial T}\right)_V =
   \left(\frac{\partial f}{\partial T}\right)_{\!\chie} \!\!
   - \left(\frac{\partial f}{\partial \chie}\right)_T
   \,\frac{(\partial \nel/\partial T)_{\chie}
   }{(\partial \nel/\partial \chie)_T},
\hspace*{1.5em}
\label{fdT}
\\&&
   \left(\frac{\partial f(\chie,T)}{\partial V}\right)_T \!\! =
   - \frac{\nel}{V}
   \left(\frac{\partial f}{\partial \nel}\right)_T \!\! =
   - \frac{\nel}{V}
   \,\frac{(\partial f/\partial \chie)_T
   }{(\partial \nel/\partial\chie)_T} .
\hspace*{1.5em}
\label{fdV}
\eea

We use analytical approximations for $I_\nu(\chie,\tau)$
based on the fits of \citet{Blin} and accurate typically to
a few parts in $10^4$, with maximum error $<0.2$\% at
$\tau\leq100$ \citep{CP98}. These approximations are given
by different expressions in three ranges of $\chie$: below,
within, and above the interval $0.6\leq\chie<14$.  In
particular, at large $\chie$ the Sommerfeld expansion (e.g.,
\citealp{Chandra,Girifalco}) yields\footnote{The 
multiplier $1/\sqrt{2}\,\tau^{\nu+1}$
was accidentally omitted in Paper~II.}
\beq
     I_\nu(\chie,\tau) =
        \frac{1}{\sqrt{2}\,\tau^{\nu+1}}\left(
        \mathcal{I}_\nu^{(0)}(\tilde\mu)
          +\frac{\pi^2}{6}\,\tau^2 \mathcal{I}_\nu^{(2)}(\tilde\mu)
            + \ldots \right),
\label{Sommer}
\eeq
where  $\tilde\mu\equiv\chie\tau=\mue/\mel c^2$ is the
electron chemical potential (without the rest energy) in the
relativistic units,
\bea&&
\mathcal{I}_{1/2}^{(0)}(\tilde\mu)
 = [\tilde{x}\tilde{\gamma}-\ln(\tilde{x}+\tilde{\gamma}) ]/2,
\label{I12}
\\&&
\mathcal{I}_{3/2}^{(0)}(\tilde\mu)
 = \tilde{x}^3/3 - \mathcal{I}_{1/2}^{(0)}(\tilde\mu),
\\&&
\mathcal{I}_{5/2}^{(0)}(\tilde\mu)
 = \tilde{x}^3\tilde{\gamma}/4 - 2 \tilde{x}^3/3
  + 1.25\,\mathcal{I}_{1/2}^{(0)}(\tilde\mu),
\label{I52}
\\&&
   \mathcal{I}_\nu^{(n+1)}(\tilde\mu)
   =
    {\dd \mathcal{I}_\nu^{(n)}}/{\dd\tilde\mu} .
\eea
Here, we have introduced notations
$\tilde{x} \equiv \sqrt{\tilde\mu(2+\tilde\mu)}$ and
$\tilde{\gamma} \equiv \sqrt{1+\tilde{x}^2} = 1+\tilde\mu$.
At strong degeneracy, $\tilde{\mu}\approx\EF/\mel c^2-1$,
$\tilde{x}\approx\xr$, and $\tilde{\gamma}\approx\gr$. In
Paper~II we also described an alternative expansion in powers
of $\tau$, which allows one to avoid numerical cancellations
of close terms at small $\tilde\mu$ 
(we switch to this alternative expansion at $\tilde\mu<0.1$). 

The discontinuities of the \citet{Blin} approximations for
$I_\nu(\chie,\tau)$ at $\chie=0.6$ and $\chie=14$ are
typically a few parts in $10^4$ at $\tau\lesssim10^2$, but
they may reach $\approx1$\% for the second derivatives. 
This accuracy is sufficient for many applications.
Nevertheless, the jumps may produce problems, e.g., when
higher derivatives are evaluated numerically in a stellar
evolution code. In our calculations of white-dwarf evolution
(to be published elsewhere), we smoothly interpolate between
the two analytical approximations for the adjacent intervals
near the boundary at the cost of a slight violation of the
thermodynamic consistency in the interpolation regions (this
version of the EOS code is now also available at our web
site).

If a higher accuracy is needed, one can numerically
calculate tables of $I_\nu(\chie,\tau)$
\citep[e.g.,][]{TimmesArnett} and interpolate in them
with an algorithm that preserves thermodynamic
consistency \citep{TimmesSwesty} and is available at
\textsc{MESA} \citep{MESA}.

\subsection{Nonideal contributions}
\label{sect:nonid}

\subsubsection{Electron and ion liquids}
\label{sect:liq-nm}

The contribution to the free energy due to the
electron-electron interactions has been studied by many authors.
For the reasons explained in Paper~II, we
adopt the fit to $F_\mathrm{ee}$ derived by \citet{IIT}
(see Appendix~\ref{sect:Fee}).

The ion-ion interactions are described using the
OCP model. In the liquid regime, the numerical results
obtained for the OCP of nonrelativistic pointlike charged
particles in different intervals of the Coulomb coupling
parameter from  $\Gamma=0$ to $\Gamma\sim200$ by different
numerical and analytical methods are reproduced by a simple
expression 
given in Appendix~\ref{sect:OCPliq}. The accurate
fit for classical OCP is supplemented by the
Wigner-Kirkwood correction, which extends the
applicability range of our approximations to lower
temperatures $T\sim\Tp$. In spite of the
significant progress in numerical \emph{ab initio} modeling
of quantum ion liquids, available results do not currently
allow us to establish an analytical extension to still lower
temperatures $T\ll\Tp$ \citep[see][for references and
discussion]{CDP02}.

\subsubsection{Coulomb crystal}

At $T< \Tm$, where $\Tm$ is the melting temperature, the ions
in thermodynamic equilibrium 
are arranged in the body-centered cubic (bcc) Coulomb lattice. 
In the harmonic approximation
(e.g., \citealp{Kittel-Q}),
the free energy of the lattice is
\beq
   F_\mathrm{lat} = U_0 + \Uq + F_\mathrm{th},
\label{f_i_harm}
\eeq
where
$U_0 = \Nion C_0 (Ze)^2/\aion$ is the classical static-lattice energy,
$C_0\approx-0.9$ is the Madelung constant, 
\beq
   \Uq=\frac32 \Nion \hbar\omp u_1
\label{uq}
\eeq
 accounts for zero-point quantum
vibrations,
$u_1=\langle\omega_{\bm{k}\alpha}\rangle_\mathrm{ph}/\omp\approx0.5$
is the reduced first moment of phonon frequencies,
\beq
   F_\mathrm{th} = 3 \Nion \kB T \left\langle\ln
   [1-\exp(-\hbar\omega_{\bm{k}\alpha} / \kB T)]
   \right\rangle_\mathrm{ph}
\eeq
is the thermal contribution, 
$\omega_{\bm{k}\alpha}$ are phonon frequencies, and
$\langle\ldots\rangle_\mathrm{ph}$ denotes the averaging
over phonon polarizations $\alpha$ and wave vectors $\bm{k}$ in
the first Brillouin zone. Here we do not separate the
classical-gas free energy, therefore $F_\mathrm{lat}$
replaces $F_\mathrm{id}^\mathrm{(i)} +
F_\mathrm{ii}^{\phantom{}}$ in \req{f-tot}. 

Beyond the harmonic-lattice approximation,
the total reduced free energy $f_\mathrm{lat}\equiv
F_\mathrm{lat}/\Nion k_B T$ can be written as
\beq
   f_\mathrm{lat} = C_0\Gamma + 1.5 u_1 \eta + f_{\mathrm{th}} +
   f_{\mathrm{ah}} .
\label{flat}
\eeq
Here, the first three terms correspond to the three terms in
\req{f_i_harm}, and $f_{\mathrm{ah}}$ is the anharmonic
correction. The most accurate values of the constants $C_0$
and $u_1$ were calculated by \citet{Baiko-disser} (see
Appendix~\ref{sect:OCPsol}). For
$f_{\mathrm{th}}=F_{\mathrm{th}}/\Nion\kB T$, we use the
highly precise fit of \citet{Baiko-ea01} 
(Appendix~\ref{sect:OCPsol}).
In the classical limit $\eta\ll1$, it reduces to
$
 f_\mathrm{th}\simeq 3\ln\eta-2.49389-1.5 u_1 \eta+\eta^2/24 ,
$
where the term with $u_1$ cancels that in \req{flat} and the
last term represents the Wigner-Kirkwood quantum
correction
$f_\mathrm{ii}^{(2)}$ [\req{fq}], which is the
same in 
the liquid and solid
phases \citep{PollockHansen}. In the opposite
limit $T\ll\Tp$, we have
$
   f_\mathrm{th} \simeq -209.3323\,\eta^{-3}
$
(here the constant is given for the bcc
crystal; for other lattice types, see
\citealp{Baiko-ea01}).

Anharmonic corrections for Coulomb lattices were studied by
many authors in the limits $\eta\to0$ and $\eta\to\infty$,
but only a few numerical results of low precision are
available at finite $\eta$ values (see Paper~II for
references and discussion). In Paper~II we constructed
an analytical interpolation between these limits, which is
applicable at arbitrary $\eta$ and is consistent with the
available numerical results within accuracy of the latter
ones
(Appendix~\ref{sect:OCPsol}).
It should be replaced by a more accurate function in
the future when accurate finite-temperature anharmonic
quantum corrections become available.

\subsubsection{Electron polarization}
\label{sect:ei}

Electron polarization in Coulomb liquids was studied by
perturbation \citep{GalamHansen,YaSha} and hypernetted-chain
(HNC)
techniques (\citealp{ChabAsh,CP98}; Paper~I). The results 
have been
reproduced by an analytical expression 
(Appendix~\ref{sect:ei-liq}), which
exactly recovers the Debye-H\"uckel limit for the
weakly coupled ($\Gamma\ll1$)
electron-ion plasmas and the Thomas-Fermi
limit for the strongly coupled ($\Gamma\gg1$) Coulomb
liquids
at $Z\gg1$.

For classical ions, the simplest screening model consists in
replacing the Coulomb potential by the Yukawa potential.
Molecular-dynamics and path-integral Monte Carlo simulations
of classical liquid and solid Yukawa systems were performed
in several works
\citep[e.g.,][]{Hama-Yukawa,MilitzerGraham}.  However, the
Yukawa interaction reflects only the small-wavenumber
asymptote of the electron dielectric function
\citep{Jancovici,GalamHansen}. The first-order perturbation
approximation for the dynamical matrix of a classical
Coulomb solid with the polarization corrections was
developed by \citet{PollockHansen}. The phonon spectrum in
such a quantum crystal has been calculated only in the
harmonic approximation \citep{Baiko02}, which has a
restricted applicability to this problem (for example, it is
obviously incapable of reproducing the polarization
contribution to the heat capacity in the classical limit
$\eta\to0$, where it gives $C_V=3\Nion\kB$ independent
of the polarization).

In Paper~I we calculated $F_\mathrm{ie}$ using the
semiclassical perturbation theory of \citet{GalamHansen}
with a model structure factor, and fit the results by an
analytical function of $\xr$ and $\eta$. In
Paper~II we improved the $\eta$-dependence of this function
to completely eliminate the screening contribution in
the strong quantum limit $\eta\ll1$, because the employed
model of the structure factor failed at $\eta \lesssim 1$.
The latter approximation is reproduced in
Appendix~\ref{sect:ei-sol}.
It can be improved in the future, when the polarization
corrections for the quantum Coulomb crystal at $\eta
\lesssim 1$ have been accurately evaluated.

\subsubsection{Ion mixtures}
\label{sect:mix}

In Sects.~\ref{sect:liq-nm}\,--\,\ref{sect:ei}
we have considered plasmas containing identical ions.
In the case where several ion types are present in a
strongly coupled Coulomb plasma,
a common approximation is the linear mixing rule
(LMR),
\beq
f_\mathrm{ex}^\mathrm{LM} 
\approx \sum_j x_j f_\mathrm{ex}(\Gamma_j,x_j=1)\,,
\label{LMR}
\eeq
where $x_j$ are the number fractions of ions with charge
numbers $Z_j$ and $\Gamma_j=\Gamme \,Z_j^{5/3}$. In
\req{LMR}, $f_\mathrm{ex}\equiv F_\mathrm{ex}/\Nion\kB T$ is
the reduced nonideal part of the free energy,
$F_\mathrm{ex}$ is the excess free energy, which is
equal to $F_\mathrm{ii}$ in the case
of the rigid
charge-neutralizing electron background and
to
$F_\mathrm{ii}+F_\mathrm{ie}+ F_\mathrm{ee}$ in the case of the
polarizable background.
The high accuracy of \req{LMR} for binary ionic mixtures in
the rigid background was first demonstrated by calculations
in the HNC approximation \citep{HTV77,BHJ79}
and confirmed later by Monte Carlo simulations
\citep{DWSC96,Rosenfeld96,DWS03}. The validity of the LMR
in the case of an ionic mixture immersed  in a
polarizable finite-temperature electron background
has been examined by \citet{HTV77} in the 
first-order thermodynamic perturbation approximation and  by
\citet{ChabAsh} by solving the HNC
equations with effective screened potentials. These authors
found that the LMR remains accurate when the electron
response is taken into account in the inter-ionic potential,
as long as the Coulomb coupling is strong ($\Gamma_j>1$,
$\forall j$).

However, the LMR is not exact, and \req{LMR}
should be replaced by the Debye\,--\,H\"uckel formula in the
limit of weak coupling ($\Gamma_j\ll1$, $\forall j$).  Even
in the strong-coupling regime, the small deviations from the
LMR are important for establishing phase equilibria
\citep[see][]{MedinCumming10}. The deviations from the LMR
were studied by
\citet{BHJ79,ChabAsh,DWSC96,DWS03,PCR09,PCCDR09} for
strongly coupled Coulomb liquids and by
\citet{Ogata-ea93} and \citet{DWS03} for Coulomb solids. 

The analytical expression that describes deviations from the
LMR, $\Delta f\equiv f-f_\mathrm{LM}$,
in Coulomb liquids for arbitrary coupling parameters
$\Gamma_j$ reads
\citep{PCCDR09}
\beq
   \Delta f_\mathrm{liq}
 = \frac{\Gamme^{3/2} \langle Z^{5/2}\rangle}{\sqrt{3}}
 \,\frac{\delta}{
 (1 + a\, \langle\Gamma\rangle^\alpha)
   \,(1+ b\,\langle\Gamma\rangle^\alpha)^\beta},
\eeq
where $\langle Z^k \rangle \equiv \sum_j x_j Z_j^k$,
$\langle\Gamma\rangle=\Gamme\langle Z^{5/3} \rangle$, 
$\delta$ is defined either as
\beq
   \delta = 1 -
    \frac{\langle Z^2\rangle^{3/2}}{\langle Z\rangle^{1/2}
          \,\langle Z^{5/2}\rangle}
\eeq
for rigid electron background model, or as
\beq
   \delta =  \frac{\langle Z\,(Z+1)^{3/2}\rangle}{\langle Z^{5/2}\rangle}
   - \frac{(\langle Z^2\rangle+\langle Z\rangle)^{3/2}
   }{\langle Z\rangle^{1/2}\,\langle Z^{5/2}\rangle}
\label{delta2}
\eeq
for polarizable background, and parameters $a$, $b$, $\alpha$, and
$\beta$ depend on the plasma composition as follows:
\bea&&
   a = \frac{2.6\,\delta+14\,\delta^3}{1-\alpha},
\quad
   \alpha = \frac{ \langle Z \rangle^{2/5}}{\langle Z^2 \rangle^{1/5} },
\label{alpha}\\&&
   b = 0.0117\,\left(\frac{\langle Z^2 \rangle }{
    \langle Z \rangle^2}\right)^{\!\!2} a\,,
\quad
   \beta = \frac{3}{2\alpha}-1.
\label{beta}
\eea

For Coulomb solids, one should distinguish regular crystals
containing different ion types and disordered solid
mixtures, where different ions are randomly distributed in
regular lattice sites \citep{Ogata-ea93}. Each regular
lattice type corresponds to a fixed composition, whereas
random lattices allow variable fractions of different ion
types. The free energy correction $\Delta f$ mainly arises
from the difference in the Madelung energies. It is
generally larger for regular crystals than for ``random''
crystals with the same composition. \citet{Ogata-ea93}
performed Monte Carlo simulations of solid ionic mixtures
and fitted the calculated deviation, $\Delta
f_\mathrm{sol}$, from linear-mixing prediction for
the reduced free energy in a random binary ion
crystal. \citet{MedinCumming10} and
\citet{Hughto-ea12} used this fit to study the
phase separation and solidification of ion mixtures in the
interiors of white dwarfs. We note, however, that the fit of
\citet{Ogata-ea93} exhibits nonphysical features: for
example, it is nonmonotonic as a function of $R_Z=Z_2/Z_1$
at a fixed number fraction $x_2=1-x_1$ for a binary ion
mixture with $Z_2/Z_1>2$ and $x_2<0.5$. A much simpler fit,
which does not exhibit unphysical behavior,
was suggested by \citet{DWS03}. It can be written as
$
\Delta f_\mathrm{sol}=0.00326\, x_1x_2 R_Z^{3/2}\langle
\Gamma \rangle.
$
However, the latter fit is
valid only for relatively small charge ratios
$R_Z\lesssim3/2$. 
We replace it by the expression
\beq
   \Delta f_\mathrm{sol} = x_1 x_2\,\Gamma_1\, \Delta g(x_2,R_Z),
\eeq
where
\beq
\Delta g(x_2,R_Z) = 0.012\,\frac{x\,(1-x)}{x_2(1-x_2)}\,
         (1-R_Z^{-2})\,(1-x_2+x_2 R_Z^{5/3})
\label{solmix}
\eeq
and
$   x = x_2/R_Z+(1-R_Z^{-1})\,x_2^{R_Z}.
$
The approximation in \req{solmix} reproduces reasonably well
the results of both \citet{Ogata-ea93} and \citet{DWS03} for
random two-component ionic bcc lattices. For a
multicomponent ion crystal, \citet{MedinCumming10} proposed
the extrapolation from the two-component plasma
case
\beq
   \Delta f_\mathrm{sol} = \sum_i \sum_{j\,>\,i} x_i x_j\,
      \Gamma_i\,
 \Delta g\left(\frac{x_j}{x_i+x_j},\frac{Z_j}{Z_i}\right),
\eeq
where the indices are arranged so that $Z_j < Z_{j+1}$.

\section{EOS of a fully ionized magnetized gas}
\label{sect:mag-id}

\subsection{Ions}

We consider only nondegenerate and nonrelativistic ions 
(for a discussion of the EOS of degenerate nuclear matter
in strong magnetic fields see, e.g., \citealp{Broderick,SuhMathews}).
In this case (cf.\ \citealp{PCS})
\beq
 \frac{F_\mathrm{id}^\mathrm{(i)}}{\Nion \kB T} =
      \ln\left(2\pi \frac{\nion \lambdi\am^2}{Z}\right)
    + \ln\left( 1- \mathrm{e}^{- \zeti}\right) -1 +
      \frac{\zeti}{2} + \frac{F_\mathrm{spin}}{\Nion \kB T} .
\label{Fp}
\eeq
The last term arises from the energy of the magnetic
moments of the ions in the magnetic field,
\beq
  F_\mathrm{spin} = - \Nion 
     \kB T \ln\Bigg[\frac{
     \sinh(\gfact\,\zeti \Mspin/4) }{ \sinh(\gfact\,\zeti/ 4)
       }
     \Bigg],
\label{F0}
\end{equation}
where $\Mspin$ is the ion spin multiplicity, 
and $\gfact$ is the $g$-factor
($\gfact=5.5857$ for
protons). 
For ions with spin one-half ($\Mspin=2$), the expression in
the square brackets in \req{F0} simplifies to
$[2\cosh(\gfact\,\zeti/ 4)]$.
For zero-spin ions, such as $^4$He,
$^{56}$Fe, and other even-even nuclei in the ground state,
$F_\mathrm{spin}=0$.

The ion pressure obeys the nonmagnetic 
ideal-gas relation $P_\mathrm{id}^\mathrm{(i)}=\nion\kB T$,
but expressions for the internal energy
and heat capacity are different:
\bea
   \frac{U_\mathrm{id}^\mathrm{(i)}}{\Nion \kB T} &=&
   \frac12 + \frac{\zeti}{e^{\zeti}-1} +
   \frac{\zeti}{2} + u_\mathrm{spin},
\\
   \frac{C_{V,\mathrm{id}}^\mathrm{(i)}}{\Nion \kB} &=&
      \frac12 + \left(\frac{\zeti}{e^{\zeti}-1}\right)^{\!2}
       + c_\mathrm{spin} .
\eea
Here, the terms $u_\mathrm{spin}$ and $c_\mathrm{spin}$ 
arise from $F_\mathrm{spin}$,
\bea
   u_\mathrm{spin} &=&
   \frac{\gfact\,\zeti/4}{\tanh(\gfact\,\zeti/4)}
    - \frac{\gfact\,\zeti \Mspin/4}{\tanh(\gfact\,\zeti \Mspin/4)},
\\
   c_{\mathrm{spin}} &=& \left(\frac{\gfact\,\zeti/4}{\sinh(\gfact\,\zeti /4)}\right)^{\!2} -
   \left(\frac{\gfact\,\zeti \Mspin/4}{\sinh(\gfact\,\zeti \Mspin/4)}\right)^{\!2} .
\eea 
They simplify at $\Mspin=2$:
\beq
   u_\mathrm{spin} =
    - \frac{\gfact\,\zeti}{4}\tanh\Bigg(\frac{\gfact\,\zeti}{4}\Bigg),
\quad
   c_{\mathrm{spin}} =
    \left(\frac{\gfact\,\zeti/4}{\cosh(\gfact\,\zeti /4)}\right)^{\!2} .
\eeq

\subsection{Electrons}
\label{sect:mag-e}

\subsubsection{General case}

Thermodynamic functions of the electron gas in a magnetic
field are easily derived from first principles
\citep{LaLi-SP1}. The number of quantum states per
longitudinal momentum interval $\Delta p_z$ for an electron
with given $\bm{B}$-projections of the spin and the orbital
moment and with a fixed Landau number $n$ in a volume $V$
equals $V\Delta p_z/(4\pi^2 a_\mathrm{m}^2 \hbar)$
\citep{LaLi-QM}. Thus one can express the electron number
density $\nel$ and the grand thermodynamic potential
$\Omega_\mathrm{id}^{(\mathrm{e})} = -P_\mathrm{id}^{(\mathrm{e})}V$ as
\bea
   \nel &=& \frac{1}{ 4\pi^2 a_\mathrm{m}^2 \hbar}
   \sum_{n=0}^{\infty} \sum_\sigma \int_{-\infty}^\infty
   \frac{\dd p_z}{\exp[(\epsilon_n(p_z)-\mue)/\kB T)]+1},
\label{n_e_mag}
\\
   \Omega_\mathrm{id}^{(\mathrm{e})} &=&
     -\frac{V \kB T}{2\pi^2 a_\mathrm{m}^2 \hbar}
         \sum_{n=0}^{\infty} \sum_\sigma
           \int_0^\infty \!\!\!
   \ln\left( 1+\exp\left[\frac{\mue-\epsilon_n(p_z)}{\kB T}\right]\right)
          \dd p_z,
\label{Omega}
\eea
where $\epsilon_n(p_z)$ is given by \req{magnenergy} and
$\sum_\sigma$ denotes the sum over spin projections, which
amounts to the factor 2 for $n\geq1$ since we neglect the
anomalous magnetic moment of electrons. This derivation
equally holds in the relativistic and nonrelativistic
theories. Equations~\ref{n_e_mag} and \ref{Omega} can be
rewritten, using integration by parts, as
\bea
   \nel = 
  \frac{1}{\pi^{3/2}\am^2\lambde}
   \sum_{n=0}^{\infty} \sum_\sigma
    (1+2bn)^{1/4}\,
     \frac{\partial I_{1/2}(\chi_n,\tau_n)}{\partial \chi_n},
\hspace*{1em}
\label{densmag}
\\
    P_\mathrm{id}^\mathrm{(e)} =
  \frac{\kB T}{\pi^{3/2}\am^2\lambde}
   \sum_{n=0}^{\infty} \sum_\sigma
    (1+2bn)^{1/4}\, I_{1/2}(\chi_n,\tau_n),
\hspace*{1em}
\label{presmag}
\eea
where
$\tau_n=\tau/\sqrt{1+2bn}
$
and
$
   \chi_n=\chie+\tau^{-1}-\tau_n^{-1}.
$
The free energy $F_\mathrm{id}^\mathrm{(e)}$ is given by
Eqs.~\ref{id_e}, \ref{densmag}, and \ref{presmag}.

The calculation of $\nel$, $P_\mathrm{id}^\mathrm{(e)}$,
and their derivatives at given $\chie$ and $\tau$ can be
performed using Eqs.~\ref{densmag}, \ref{presmag} and
the same analytical approximations to the Fermi-Dirac
integrals as for the nonmagnetized electron gas. The reduced
electron chemical potential $\chie$ at constant $\nel$ and
$T$ is found by numerical inversion of \req{densmag}.
Then the derivatives over $T$ at constant $V$ and over
$V$ at constant $T$ are given by Eqs.~\ref{fdT} and
\ref{fdV}. We use this approach in the current research,
but we should note that for quantizing magnetic fields it is
less precise than at $B=0$. As mentioned in
Sect.~\ref{sect:id}, the inaccuracy of the employed
approximations for $I_\nu(\chie,\tau)$ is within a fraction
of percent, but it grows for the derivatives. Since the
first derivatives are already employed in \req{densmag},
evaluation of the second-order thermodynamic functions such
as $\chi_T$ or $C_V$ involves third derivatives. Therefore,
the error in the evaluation of these functions may rise to
several percent. This level of accuracy may be sufficient for
astrophysical applications, but otherwise one should resort
to a thermodynamically consistent interpolation in numerical
tables of the Fermi-Dirac integrals
\citep{TimmesArnett,TimmesSwesty}.
Equations \ref{densmag} and \ref{presmag} can be
simplified in several limiting cases considered
below. 

\subsubsection{Strongly quantizing and nonquantizing
magnetic fields}

The field is called strongly quantizing if most of the
electrons reside on the ground Landau level. The
electron Fermi momentum, then, equals
\beq
  \pF = 2\pi^2\am^2\hbar \nel
    = (3\pi^2/2)^{1/3}\,(\am/\ael)^2\,\pF^{(0)},
\label{pFmag}
\eeq
where $\pF^{(0)}$ is the zero-field
Fermi momentum at the given density.
Equation \ref{pFmag} can be written as
$\pF =\mel c x_B$, where
\beq
   x_B = \frac{2\xr^3}{3b}
        \approx 30.2\,\frac{{Z}}{A}
               \,\frac{\rho_6}{ B_{12}}
\label{pF-mag}
\end{equation}
is the relativity parameter modified by the field and
$\xr=\pF^{(0)}/\mel c$ (Sect.~\ref{sect:gen}).
With increasing $\nel$ at constant $B$ and zero temperature,
the electrons start to populate the first excited Landau
level when $\nel$ reaches $n_B=(\pi^2\sqrt2\,\am^3)^{-1}$.
Therefore, the field is strongly quantizing at 
$T\ll T_\mathrm{cycl}$ and $\rho<\rho_B$, where
$
    T_\mathrm{cycl} = \hbar\omc/\kB
      \approx 1.3434\times10^8 B_{12}
$~K
and
\beq
    \rho_B = \mion n_B/Z
  \approx 7045 \,(A/Z)
       \,B_{12}^{3/2}\mbox{ \gcc}.
\label{rho_B}
\eeq
The condition $\nel<n_B$ can be written as $\am/\ael <
\sqrt2\,(3\pi)^{-1/3}$. Then \req{pFmag} shows that in a
strongly quantizing field $\pF < \pF^{(0)}$, except for
densities $\nel$ close to the threshold $n_B$. Thus $\TF$ is
reduced, compared to its nonmagnetic value $\TF^{(0)}$, by
factor $\TF/\TF^{(0)}=[(1+x_B^2)^{1/2}-1]/(\gr-1)$. In the
nonrelativistic limit,  $\TF/\TF^{(0)} = (\pF/\pF^{(0)})^2$,
and the parameter $\theta=T/\TF$ becomes
\beq
   \theta_B=8\gammam^2
       \rs^5/(9\pi^2\Gamme)
        \approx 0.166\,\theta_0\,\gammam^2 \rs^4,
\label{theta_B}
\eeq
where $\theta_0$ is the nonmagnetic value given by \req{theta}.

The opposite case of a nonquantizing magnetic field occurs
at $T\gg T_B$, where $T_B$ is the temperature at which the
thermal kinetic energy of the electrons becomes sufficient
to smear their distribution over many Landau levels. It can
be estimated as $T_B = T_\mathrm{cycl}$, if $\rho \leq
\rho_B$ and $T_B=T_\mathrm{cycl}/\gr$, if $\rho > \rho_B$
(a more sophisticated but qualitatively similar definition
of $T_B$ was introduced by \citealp{Lai01}).
Then we can approximately replace the sum over Landau level
numbers $n$  by the integral over a continuous variable $n$.
Integrating over $n$ by parts, we can reduce \req{presmag}
to \req{P_e} and \req{densmag} to \req{n_e}, i.e., to
recover the zero-field thermodynamics. At $\rho\gg\rho_B$,
the electrons also fill many Landau levels and the magnetic
field can be approximately treated as nonquantizing.

In the intermediate region, where the magnetic field is
neither strongly quantizing nor nonquantizing, the summation
over $n$ manifests itself in quantum oscillations of the
thermodynamic functions with changing $B$ and/or $\rho$,
similar to the de Haas -- van Alphen oscillations of magnetic
susceptibility (e.g., \citealt{LaLi-SP1}). The oscillations
are smoothed by the thermal broadening of the Fermi
distribution function and by the quantum broadening of the
Landau levels (particularly, owing to electron collisions;
see \citealt{YakovlevKaminker}, for references). Some
examples of such oscillations will be given in
Sect.~\ref{sect:examples}.

\begin{figure}
\includegraphics[width=\columnwidth]{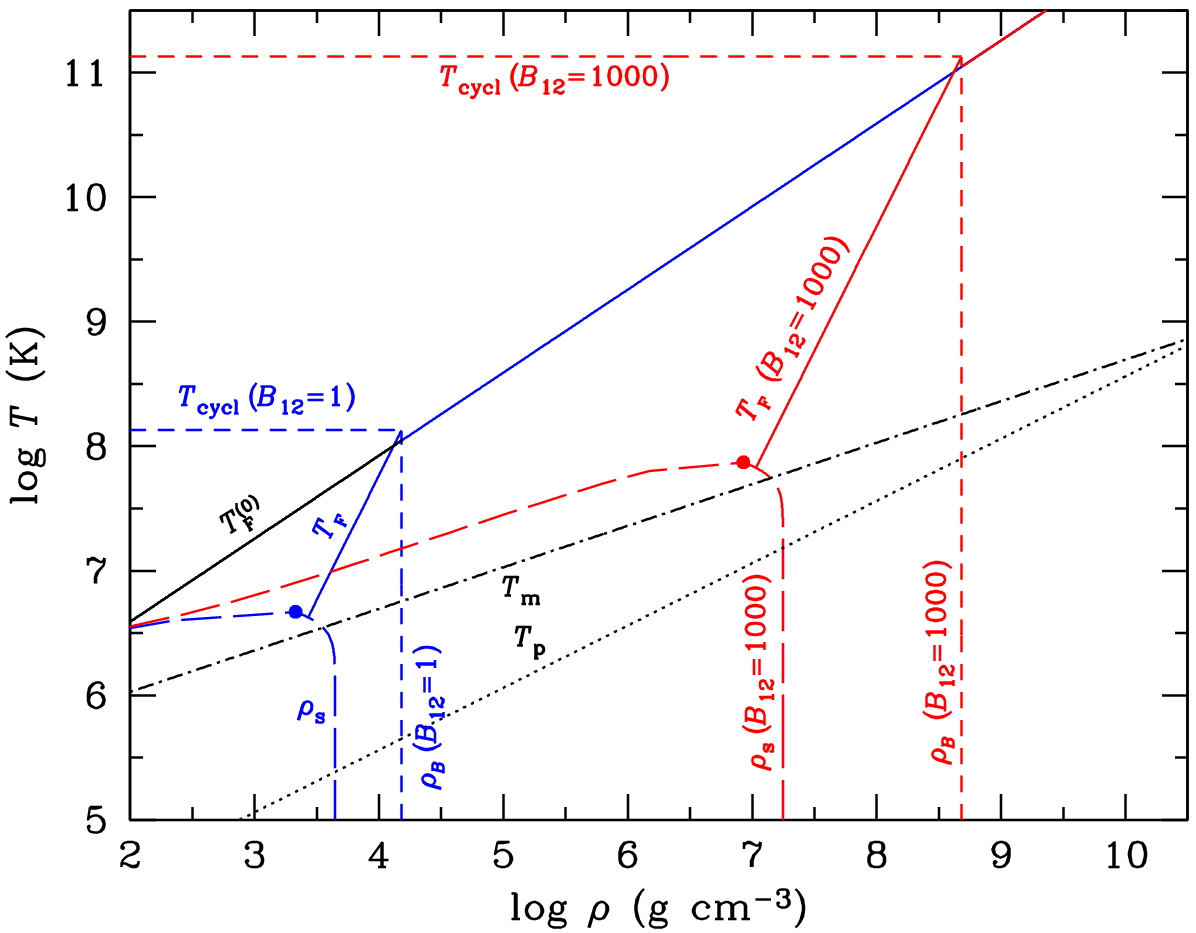}
\caption{
Characteristic density-temperature domains at $B=10^{12}$~G
(blue online) and $10^{15}$~G (red online) for fully-ionized
iron. Solid lines indicate the Fermi temperature as function
of density, the dotted line shows the plasma temperature,
the dot-dashed line shows the melting temperature as
function of density, short and long dashes delimit the
domains of strongly quantizing magnetic field and of magnetic
condensation, respectively, and the heavy dots mark the critical
point for the condensation (Sect.~\ref{sect:cond}).
}
\label{fig:diamag}
\end{figure}

Figure~\ref{fig:diamag} presents the $\rho$\,--\,$T$ diagram
of outer neutron-star envelopes at two magnetic field
strengths, $B=10^{12}$~G and $10^{15}$~G, assuming
fully-ionized iron (this assumption may be crude in the
lower left part of the diagram).  In the strongly-quantizing
magnetic-field domain, bounded by $\rho_B$ and
$T_\mathrm{cycl}$, the dependence $\TF(\rho)$ is steeper
than at $B=0$, in agreement with \req{theta_B}. The line
$\Tm(\rho)$ separates Coulomb liquid from Coulomb crystal. 
Near the lower right corner of the figure, where $T\ll\Tp$,
the quantum effects on the ions become important (i.e., 
the ions cannot be treated as classical pointlike particles).
In the lower left corner, at $\rho<\rhos$, the plasma is
unstable to the phase separation into the gaseous and
condensed phases (this phase transition will be discussed in
Sect.~\ref{sect:cond}).

\subsubsection{Strongly degenerate electrons} 

If the electrons are strongly degenerate, then one can apply
the Sommerfeld expansion (Sect.~\ref{sect:id}) and obtain
$F_\mathrm{id}^\mathrm{(e)} \approx F_0^\mathrm{(e)} +\Delta
F$, where $F_0^\mathrm{(e)} = \EF \Nel -P_0^\mathrm{(e)} V$
is the zero-temperature value and $\Delta F$ is a thermal
correction. According to Eqs.~\ref{Sommer} and
\ref{presmag}, the zero-temperature pressure $P_0$ is
\beq
   P_0^\mathrm{(e)} = \frac{\Prel\, b}{2\pi^2}
      \sum_{n=0}^{n_\mathrm{max}} \sum_\sigma
       (1+2bn)\,
      \mathcal{I}_{1/2}^{(0)}(\tilde\epsilon_n),
\eeq
where $\Prel \equiv \mel^4 c^5/\hbar^3 = 1.4218\times 
10^{25}\mathrm{~dyn~cm}^{-2}$ is the relativistic unit of
pressure, $n_\mathrm{max}$ is the maximum integer $n$ for
which $p_n^2(\epsilon)>0$, and
$\tilde\epsilon_n\equiv\EF/\mel c^2+1-\sqrt{1+2bn}$.
According to Eqs.~\ref{Sommer} and \ref{densmag}, the
Fermi energy $\EF$ is determined by the condition
\beq
   \nel = \left(\frac{\mel c}{\hbar}\right)^{3}
   \frac{b}{2\pi^2} \sum_{n=0}^{n_\mathrm{max}} \sum_\sigma
   \,(1+2bn)^{1/2}
    \mathcal{I}_{1/2}^{(1)}(\tilde\epsilon_n).
\label{n_e_deg}
\eeq
In order to obtain the chemical potential
$
   \mue = \EF + \Delta\epsilon
$
with fractional accuracy $\sim\chie^{-2}$, 
we retain two terms
in \req{Sommer}, 
insert it into \req{densmag},
approximate
$\mathcal{I}_\nu^{(n)}(\tilde\mu_n)$
in the vicinity of
$\tilde\mu_n=\epsilon_n$ by
\beq
\mathcal{I}_\nu^{(n)}(\tilde\mu_n)\approx
\mathcal{I}_\nu^{(n)}(\tilde\epsilon_n)
+\mathcal{I}_\nu^{(n+1)}(\tilde\epsilon_n)\,\Delta\tilde\epsilon ,
\label{Inu1}
\eeq
where $\tilde\mu_n\equiv\chi_n\tau$
and $\Delta\tilde\epsilon\equiv\Delta\epsilon/\mel c^2$,
and drop the higher-order terms containing
$(\tau^2\Delta\tilde\epsilon)$. Then
\beq
  \Delta\tilde\epsilon
   \approx
   - \frac{\pi^2\tau^2}{6}
   \frac{\sum_{n=0}^{n_\mathrm{max}} \sum_\sigma
     (1+2bn)^{-1/2}\mathcal{I}_{1/2}^{(3)}(\tilde\epsilon_n)
   }{\sum_{n=0}^{n_\mathrm{max}} \sum_\sigma
     (1+2bn)^{1/2}\mathcal{I}_{1/2}^{(2)}(\tilde\epsilon_n)
   } \,.
\label{DeltaE}
\eeq
The thermal correction to the pressure equals
\beq
   \Delta P = \Prel\, b
     \sum_{n=0}^{n_\mathrm{max}} \sum_\sigma
      (1+2bn) \left(
        \frac{\tau^2}{12}\,\mathcal{I}_{1/2}^{(2)}(\tilde\epsilon_n)
     + \frac{\Delta\tilde\epsilon}{2\pi^2}
      \,\mathcal{I}_{1/2}^{(1)}(\tilde\epsilon_n)
         \right)\,,
\label{DeltaP}
\eeq
and the thermal correction to the free energy and internal
energy
\beq
 \Delta F = - \Delta U
    = \Nel\, \Delta\epsilon - V \Delta P.
\label{DeltaU}
\eeq
The leading contribution to the heat capacity 
is $C_V^\mathrm{(e)} = 2\Delta U/T$. As
in the nonmagnetic case, $C_V^\mathrm{(e)}$ is
proportional to $T$ at $T\to0$, but with a different
proportionality coefficient.

\subsubsection{Strongly degenerate electrons in a strongly
quantizing magnetic field} 

If the magnetic field is strongly quantizing
and the electrons are strongly degenerate
(which corresponds to the triangular domains in
Fig.~\ref{fig:diamag} defined by conditions $\rho<\rho_B$
and $T<\TF$),
then
\bea
   F_0^\mathrm{(e)} &=& \left[(1+x_B^2)^{1/2} - 1\right]
         \Nel\mel c^2 - P_0^\mathrm{(e)} V,
\\
   P_0^\mathrm{(e)} &=& \frac{\Prel\,b}{(2\pi)^2}\,
        \left[
          x_B\,(1+x_B^2)^{1/2}
           - \ln\left(x_B+(1+x_B^2)^{1/2}\right)
            \right] .
\label{P-mag-simple}
\eea
In the nonrelativistic ($x_B\ll1$) and ultrarelativistic
($x_B\gg1$) limits, we have $P_0^\mathrm{(e)}\simeq\Prel\,
b\, x_B^3/6\pi^2 \propto \nel^3$ and
$P_0^\mathrm{(e)}\simeq\Prel\, b\, x_B^2/4\pi^2 \propto
\nel^2$, respectively. Compared with the nonmagnetic 
case (Papers~I and II), the dependence $P_0^\mathrm{(e)}(\nel)$
is steeper, but $P$ is lower everywhere except  for
$\nel\approx n_B$. Thus, a strongly quantizing
magnetic field softens the EOS of degenerate electrons.

The thermal corrections (Eqs.~\ref{DeltaE}\,--\,\ref{DeltaU})
simplify to
\bea&&
\Delta\tilde\epsilon
\approx
- \frac{\pi^2\tau^2}{6(1+x_B^2)^{1/2} x_B^2},
\\&&
\Delta P = \Prel \, \frac{b \tau^2}{12}\,
 \frac{2+x_B^2}{(1+x_B^2)^{1/2}\, x_B^{\phantom{2}}},
\\&&
\frac{\Delta U}{V} = -\frac{\Delta F}{V}
 = \Prel \,\frac{b \tau^2}{12}\, \frac{(1+x_B^2)^{1/2}}{x_B},
\\&&
\frac{C_V^\mathrm{(e)} }{ \Nel\kB} =
\frac{\pi^2\tau}{3}\,\frac{(1+x_B^2)^{1/2}}{x_B^2}\,.
\eea
The last
equation differs from the nonmagnetic case (Paper~II)
in that $x_B$ replaces $\xr$ and $\pi^2/3$ replaces
$\pi^2$.

\subsubsection{Nonrelativistic limit}

In the nonrelativistic limit ($\pF\ll \mel c$ and
$T\ll T_\mathrm{r}$), Eqs.~{\ref{densmag} and
\ref{presmag} simplify to
\beq
   \nel = \frac{1}{2\pi^{3/2} a_\mathrm{m}^2 \lambde}
     \sum_{n,\sigma} I_{-1/2}(\chi_n),
~~
   P_\mathrm{e} = \frac{\kB T}{\pi^{3/2} a_\mathrm{m}^2\lambde}
     \sum_{n,\sigma} I_{1/2}(\chi_n).
\label{n_e_mag_NR}
\end{equation}
In the nondegenerate regime ($T\gg T_\mathrm{F}$),
one has $I_\nu(\chi)\approx \mathrm{e}^{\chi}\,\Gamma(\nu+1)$,
where $\Gamma(\nu+1)$ is the gamma-function.
Then \req{n_e_mag_NR}
yields $P_\mathrm{id}^\mathrm{e} = \nel \kB T$
and
\beq
   \chie =  \ln(\nel \lambde^3/2) - \ln (\zeta/4) +
                   \ln[\tanh (\zeta/4)]
\label{chi_mag}
\eeq
which provides the free energy
$F_\mathrm{id}^{(e)} =(\chie - 1)\,\Nel\kB T$. 

As follows from \req{chi_mag}, the reduced internal energy
$U_\mathrm{id}^\mathrm{e}/\Nion\kB T$ and heat capacity
$C_{V,\mathrm{id}}^\mathrm{e}/\Nion\kB$ of the Boltzmann gas
decrease with increasing $\zeta$. In a strongly quantizing
magnetic field ($\zeta\gg1$), they tend to $1/2$  instead of
$3/2$ because the gas becomes effectively one-dimensional.
The only kinetic degree is along the magnetic
field.

In the nonquantizing field ($\zeta\ll1$), the two last terms
in \req{chi_mag} cancel out, so that the standard expression
$F_\mathrm{id}^{(e)} = \Nel\kB T \,[\ln(\nel \lambde^3/2)
-1]$ is recovered. In the strongly quantizing, nondegenerate
regime ($\rho<\rho_B$ and $\TF\ll T\ll T_\mathrm{cycl}$),
the last term of \req{chi_mag} vanishes, which yields
\beq
   F_\mathrm{id}^{(e)} =
      \Nel \kB T \left[ \ln(2\pi a_\mathrm{m}^2\lambde \nel)
         - 1 \right].
\label{Fe-magn}
\end{equation}

\subsection{Thermodynamic and kinetic pressures}

The above expressions for pressure of a magnetized gas of
charged particles are based on the principles of
thermodynamics \citep{LaLi-SP1}, according to which
$P=-(\partial F/\partial V)_{T,B}$. Alternatively, the pressure
can be calculated from the microscopic dynamics as the sum
of the changes of kinetic momenta of all particles reflected
off a unit surface per unit time. The result of the latter
calculation, the kinetic pressure $P^\mathrm{kin}$,
depends on the orientation of the surface relative to 
$\bm{B}$ \citep{CanutoChiu}. If the surface is perpendicular
to $\bm{B}$, then one gets the kinetic pressure
$P_\|^\mathrm{kin}=P$, which acts along the field lines, the
longitudinal pressure. If the
surface is parallel to the field, one gets a different
(transverse) kinetic pressure,  which can be expressed
\citep{BlandfordHernquist} as 
\beq
   P_\perp^\mathrm{kin} =
       -\Omega/V + B\,(\partial\Omega/\partial B)_{V,T}
         = P-M B, 
\eeq
where
$
   M=-\partial\Omega/\partial B
$
is the magnetization. 

In order to resolve the apparent paradox, one should take the
magnetization current density $\bm{j}_\mathrm{m}=c\,\nabla
\times {\bm{M}}$ into account; when boundaries are present,
this volume current
should be supplemented by the surface current
$c\bm{M}\times\bm{B}/B$ (see, e.g., \citealp{Griffith}).  As
argued by \citet{BlandfordHernquist}, if we compress the
electron gas perpendicular to $\bm{B}$ then we must do work
against the Lorentz force density
$\bm{j}_\mathrm{m}\times\bm{B}/c$, which gives an additional
contribution to the total transverse pressure and makes it
equal to $P$. Because this point still causes confusion in
some publications, let us illustrate it with a simple
example.

If the  pressure were anisotropic, then one might expect an
anisotropic density gradient in a strongly magnetized star.
Let us consider a small volume element in the star, assuming
that we can treat $\bm{B}$, $T$, and gravitational
acceleration $\bm{g}$ as constants within this volume, and
$z$ axis is directed along $\bm{g}$. Hydrostatic equilibrium
implies that the density of gravitational force, $\rho
\bm{g}$, is balanced by the density of forces created by
plasma particles. The crucial point is that the
magnetization contributes to this balance.

Let us compare the cases where $\bm{B}$ is parallel and
perpendicular to $\bm{g}$. In the first case, the
$z$-component of the Lorentz force is absent, and we get the
standard equation of hydrostatic equilibrium: $\rho g=\dd
P_\|^\mathrm{kin}/\dd z=\dd P/\dd z$. In the second case,
the gradients of the kinetic pressure $\dd
P_\perp^\mathrm{kin}/\dd z$  and of the Lorentz force
density $B\,\dd M/\dd z$ act in parallel. In the constant
and uniform magnetic field, $\dd M/\dd z$ is not zero, but
is related to the density gradient:
\beq
 \frac{\dd M}{\dd z}=\frac{\partial M(\rho,T,B)}{\partial\rho}\,
    \frac{\dd \rho}{\dd z}
    = - \frac{\partial^2\Omega(\rho,T,B)
       }{ V \partial\rho\, \partial B}\,
    \frac{\dd \rho}{\dd z} .
\label{dMdz}
\eeq
Then the equilibrium condition takes the form
$
 \rho g = {\dd P_\perp^\mathrm{kin}}/{\dd z}
    + B\,{\dd M}/{\dd z} ={\dd P}/{\dd z},
$
which is the same as in the first case. Furthermore, one can
express $P$ through $\rho$ using some EOS. In the considered
example, $\dd P/\dd z = [\partial
P(\rho,T,B)/\partial\rho]\,\dd \rho/\dd z$, so that $\dd
\rho/\dd z$ is also the same in both cases. Thus, the
stellar hydrostatic profile is determined by the isotropic
thermodynamic pressure $P$, which automatically includes
magnetization. 

\section{Magnetic effects on the EOS of a Coulomb liquid}
\label{sect:mag-other}

\subsection{Electron exchange and correlation}
\label{sect:ee-mag}

The effects of a magnetic field on the contribution to the
free energy due to electron exchange and correlation were
studied either in the regime of strong degeneracy and
strongly quantizing magnetic fields
(\citealp{DanzGlasser,BanerjeeCR,Fushiki-ea}; see also
\citealp{MorbecCapelle08} for an instructive discussion of
the previous results and the inclusion of the second Landau
level contribution), or at low densities
\citep{AlaJanco,Cornu,Steinberg,Steinberg2}. In a previous
work \citep{PCS} we suggested a modification of the
field-free expression for $F_\mathrm{ee}$, which matches
available exact limiting expressions, including the cases of
nonquantizing, strongly quantizing degenerate,
and strongly quantizing nondegenerate plasmas. The
modification consists in replacing
$F_\mathrm{ee}(\theta,\Gamme)$
by
$F_\mathrm{ee}(\theta^\ast,\Gamme)$,
where
\beq
  \theta^\ast =
   \frac{ \theta_0 + \theta_B 
      }{\displaystyle
          1+\frac{\theta_B}{\theta_0 }\,\exp(-\theta_B^{-1})\,
            \frac{\cosh (\zeta/2)}{[ \cosh(\zeta/4)]^2}
          \, \frac{\tanh (\zeta/4) }{ \zeta/4 } \,
          \frac{\mathrm{arctanh}\,\xi
        }{ \xi}
           }\,,
\label{theta-ast}
\eeq
$\xi=[1-(4/\zeta)\tanh (\zeta/4)]^{1/2}$,
and $\theta_0$ and $\theta_B$ are given by
Eqs.~\ref{theta} and \ref{theta_B}, respectively,
at fixed $\nel$ and $T$.

\subsection{Wigner-Kirkwood term}
\label{sect:fq-mag}

For the same reason as in Sect.~\ref{sect:liq-nm}, the
treatment of the quantum effects in the ion liquid is
restricted by the Wigner-Kirkwood term. Its expression in an
arbitrary magnetic field was obtained by \citet{AlaJanco}:
\beq
   f_\mathrm{ii}^{(2)}=\frac{\eta^2}{24}\,
    \left[ \frac{4}{\zeti\,\tanh (\zeti/2)} - \frac{8}{\zeti^2}
          + \frac{1}{3} \right].
\label{fq-mag}
\eeq
The function in the square brackets monotonously varies from
1 at $\zeti\to0$ to $1/3$ at $\zeti\to\infty$, reflecting the
effective reduction of the degrees of freedom of the
classical ion motion from $d=3$ at $B=0$ to $d=1$ for
a strongly quantizing field. At small $\zeti$,
$f_\mathrm{ii}^{(2)} \approx (\eta^2/24)\,(1-\zeti^2/90)$.

\subsection{Electron-ion correlations}

Using the linear response theory in the Thomas-Fermi limit,
\citet{Fushiki-ea} evaluated the electron polarization
energy  for a dense plasma in a strongly quantizing magnetic
field at zero temperature, assuming that the ions remain
classical (unaffected by the field). A comparison with the
analogous zero-field result shows that the strongly
quantizing magnetic field ($\gammam \rs^2 > 2.23$)
increases the polarization energy at high densities
($\rs\ll1$) by a factor of $0.8846\,\gammam^2
\rs^4$ \citep{PCS}.

Recently, \citet{SharmaReddy11} calculated the
screening of the ion-ion potential due to electrons in a
large magnetic field $B$ at $T=0$, using the one-loop
representation of the polarization function. Their results
for the strongly quantizing magnetic field show that the
screening is anisotropic, and the screened ion potential
exhibits Friedel oscillations with period $\pi\hbar/\pF$ in
a cylinder of a radius $\sim\pi\hbar/\pF$ along the magnetic
field line that passes through the Coulomb center.
\citeauthor{SharmaReddy11} suggest that this long-range
oscillatory behavior can affect the ion lattice structure.
However, finite temperature should damp these oscillations,
so that they are pronounced only at $T\ll\TF$, i.e., deep
within the triangular domains formed by the lines $\TF$ and
$\rho_B$ in Fig.~\ref{fig:diamag}. At the typical pulsar
magnetic fields $B\sim10^{12}$~G, this requires an unusually
low temperature of the neutron-star crust. On the other
hand, the conditions  $T\ll\TF$ and $\rho<\rho_B$ can be
easily fulfilled in the outer crust of magnetars at
$B\sim10^{15}$~G (cf.~Fig.~\ref{fig:diamag} and the top
panel of Fig.~\ref{fig:eospw}), but in this case the Friedel
oscillations are strongly suppressed because the electrons
are ultrarelativistic.

To the best of our knowledge,
the magnetic effects on the
electron polarization energy have not been calculated
at finite temperatures or in the case where the field is
not strongly quantizing.
In view of the limited scope and limited applicability of
the available results on the magnetic effects, we use the
nonmagnetic expression for $F_\mathrm{ie}$ in our code
(Appendix~\ref{sect:ei2}).

\section{Harmonic Coulomb crystals in the magnetic field}
\label{sect:mag-sol}

The magnetic effects on Coulomb crystals have been studied
only in the harmonic approximation.
\citet{NagaiFukuyama,NagaiFukuyama2} calculated phonon
spectra of body-centered cubic (bcc), face-centered cubic
(fcc), and hexagonal closely-packed (hcp) OCP lattices. They
compared the energies of zero-point vibrations at different
values of parameters $\beta$ and $\RS$ and found
conditions of stability of every lattice type. However,
\citet{Baiko-disser,Baiko09} noticed that their choice of
the magnetic-field direction did not provide the minimum of
the total energy. 

\citet{UsovGU} obtained the equations for oscillation modes
of a harmonic OCP crystal and studied its phonon spectrum in
a quantizing magnetic field in several limiting cases. 
These authors discovered a ``soft'' phonon mode with
dispersion relation $\omega_{\bm{k}\alpha}\propto k^2$ near
the center of the Brillouin zone, which leads to the
unusual  dependence of the heat capacity of the lattice
$C_{V,\mathrm{lat}}\propto T^{3/2}$ at  $T\to0$ instead of
the Debye law $C_{V,\mathrm{lat}}\propto T^3$.
\citet{UsovGU}  argued that a strong magnetic field should
increase stability of the crystal.

\citet{Baiko-disser,Baiko09} studied the magnetic effects on
the phonon spectrum of the harmonic Coulomb crystals and
calculated its energy, entropy, and heat capacity. We have
found that his results can be approximately reproduced by
the analytical expressions presented below.

\subsection{Thermal phonon contributions}

\begin{figure}
\includegraphics[width=\columnwidth]{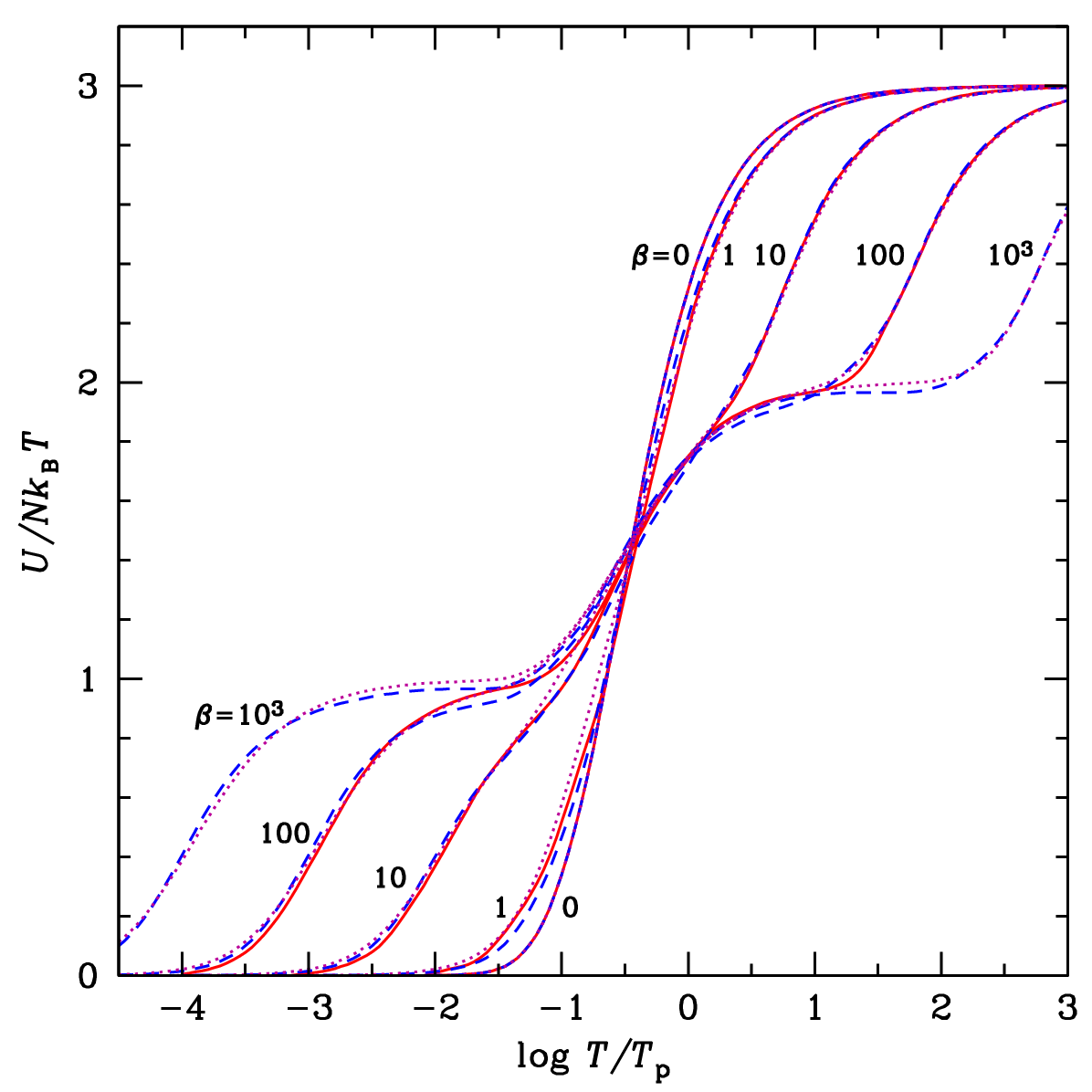}
\caption{
Thermal phonon contribution to the reduced internal energy
$u_{\mathrm{th}}=U_{\mathrm{th}}/\Nion\kB T$ as a function
of $\log(T/\Tp)=-\log\eta$ at $\beta=\hbar\omci/\kB\Tp = 0$, 1,
10, 100, and $10^{3}$ (numbers near the lines).
The analytical approximation in \req{ufit} (dotted lines) and
in \req{ufit1} (short-dashed lines) are compared with the
numerical results of \citet{Baiko09} (solid lines for
$\beta=1$, 10, and 100).
}
\label{fig:uth}
\end{figure}

Without a magnetic field, the thermal phonon contribution
$f_{\mathrm{th}}$ to the reduced free energy of a Coulomb
crystal $F/\Nion\kB T$ is a function of a single argument
$\eta$, described by a simple analytical expression
\citep{Baiko-ea01}. The magnetic field introduces the second
independent dimensionless argument $\beta$. The functional
dependence of thermodynamic functions on $\eta$ and $\beta$
is not simple. \citet{Baiko09} identified five
characteristic sectors of the $\eta$\,--\,$\beta$ plane: 
\\
\hspace*{1em}1.~~$\eta < 1$  and  $\beta < \eta^{-1}$
 -- weakly magnetized classical crystal, \\
\hspace*{1em}2.~~$\eta > 1$  and  $\beta < \eta^{-1}$
 -- weakly magnetized quantum crystal, \\
\hspace*{1em}3.~~$\eta < 1$  and  $\beta > \eta^{-1}$
 -- strongly magnetized classical crystal, \\
\hspace*{1em}4.~~$\eta > \beta > \eta^{-1}$
 -- strongly magnetized quantum crystal, \\
\hspace*{1em}5.~~$\beta > \eta > 1$
 -- very strongly magnetized quantum crystal.
\\
Note that the condition $\beta > \eta^{-1}$ is equivalent to
$\zeti > 1$. Thus, the magnetic field strongly affects the
thermodynamic functions of a Coulomb crystal when
$\hbar\omci > \kB T$, that is the same condition as for the
gas and liquid phases.

\begin{figure}
\includegraphics[width=\columnwidth]{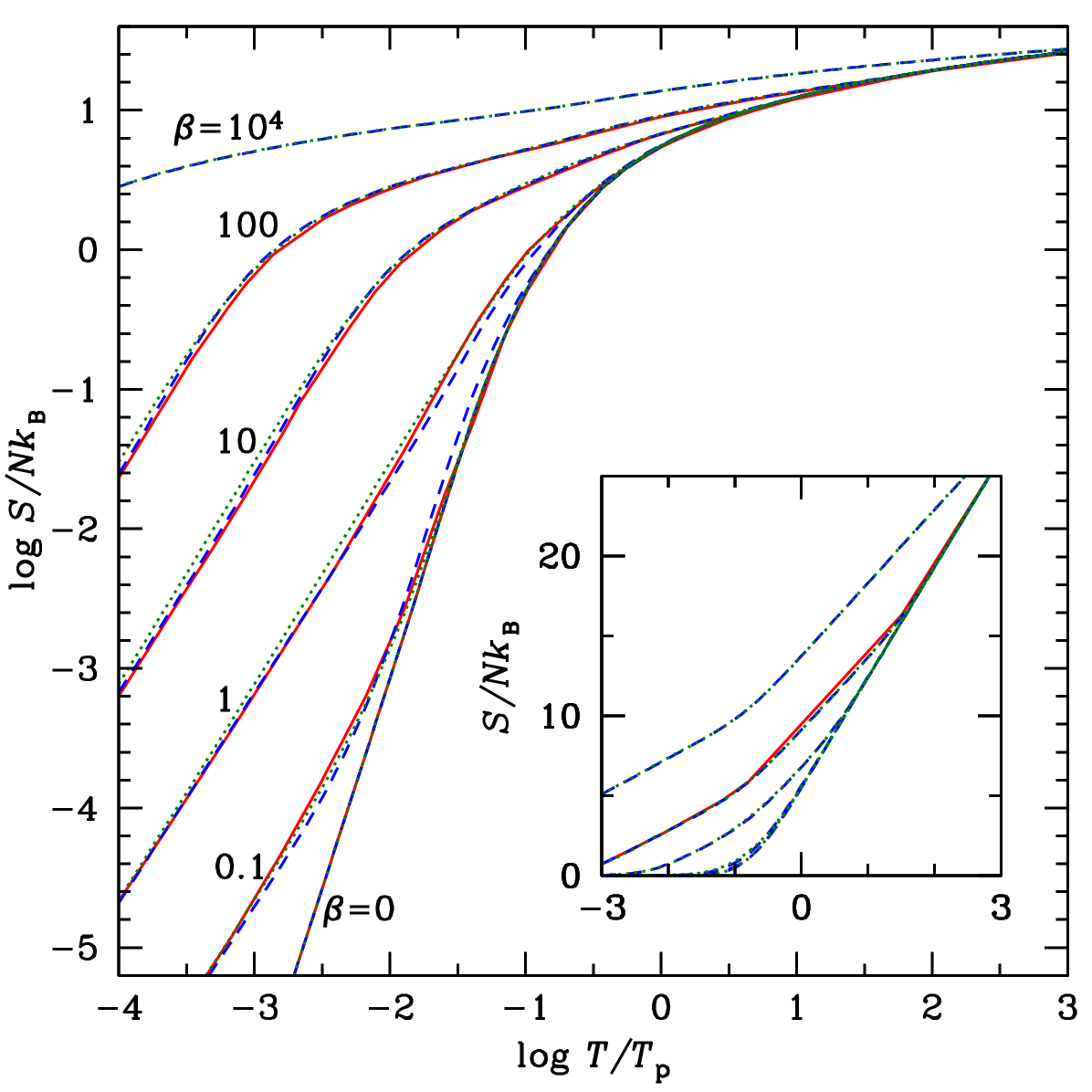}
\caption{
Thermal phonon contribution to the reduced entropy
$s_{\mathrm{th}}=S_{\mathrm{th}}/\Nion\kB T$ as a function
of $\log(T/\Tp)$ at $\beta=\hbar\omci/\kB\Tp = 0$, 0.1, 1,
10, 100, and $10^{4}$ (numbers near the lines).
The analytical approximations in \req{sfit} (dotted lines)
and in \req{sfit1} (dashed lines) are compared with the
numerical results of \citet{Baiko09} (solid lines for
$\beta\leqslant100$). The
inset shows the same approximations in the linear scale for
$s_{\mathrm{th}}$; here, the solid line corresponds
to the numerical data at $\beta=100$.
}
\label{fig:sth}
\end{figure}

For astrophysical applications, we have constructed
an analytical representation of the EOS of the magnetized
Coulomb crystal, which is asymptotically exact in each of the
five sectors far from their boundaries, exactly recovers the
nonmagnetic fit of \citet{Baiko-ea01} in the limit
$\beta\to0$, and reaches a reasonable compromise between
simplicity and accuracy.

The term
$f_{\mathrm{th}}$ in \req{flat} can be rewritten as
$
   f_{\mathrm{th}}=u_{\mathrm{th}}-s_{\mathrm{th}},
$
where
$u_{\mathrm{th}}=U_{\mathrm{th}}/\Nion\kB T$
and $s_{\mathrm{th}}=S_{\mathrm{th}}/\Nion\kB$
are the thermal
contributions to the reduced internal energy and entropy.
We approximately represent $u_{\mathrm{th}}$ by the
function
\beq
     \tilde{u} =
        \frac{u_{\mathrm{th}}^{(0)}
             }{
             1+0.5\left[1+(3/\zeti)^{5/2}\right]^{-2/5}}
        + \frac{\psi\,/\sqrt{1+24/\eta^2}
              }{
               1+15\,\beta/\eta+\psi}     ,
\label{ufit}
\eeq
where $\psi = 12.5\,(\beta/\eta)^{3/2}
               + 119\,(\beta/\eta)^2$
and $\zeti=\beta\eta$,
and we represent $s_{\mathrm{th}}$ by the
function
\beq
     \tilde{s} = s_{\mathrm{th}}^{(0)} +
        \ln\left(1+
          \frac{20.8\,(\beta/\eta)^{3/2}
             + 122.5\,(\beta/\eta)^2
                 }{
               (1 + 7.9\,\beta/\eta)\,
               (1 + 42/\eta^2)\,
               (1 + \zeti^{-1})^4}
               \right).
\label{sfit}
\eeq
In these equations, $u_{\mathrm{th}}^{(0)}$ and
$s_{\mathrm{th}}^{(0)}$ are the values of $u_{\mathrm{th}}$
and $s_{\mathrm{th}}$ at
$\beta=0$.
Equations (\ref{ufit}) and (\ref{sfit}) exactly reproduce
the known asymptotic limits: $u_{\mathrm{th}}=3$
in the classical nonmagnetic limit ($\eta\ll\beta\ll1$),
$u_{\mathrm{th}}=2$
in the classical magnetic limit ($\eta\ll\beta^{-1}\ll1$),
$u_{\mathrm{th}}=1$ in the case where 
$\beta\gg\eta\gg1$,
and $u_{\mathrm{th}} = 0.6 s_{\mathrm{th}} \propto
\eta^{-3/2}$, if $\eta\to\infty$ at
$\beta={}$constant.

\begin{figure}
\includegraphics[width=\columnwidth]{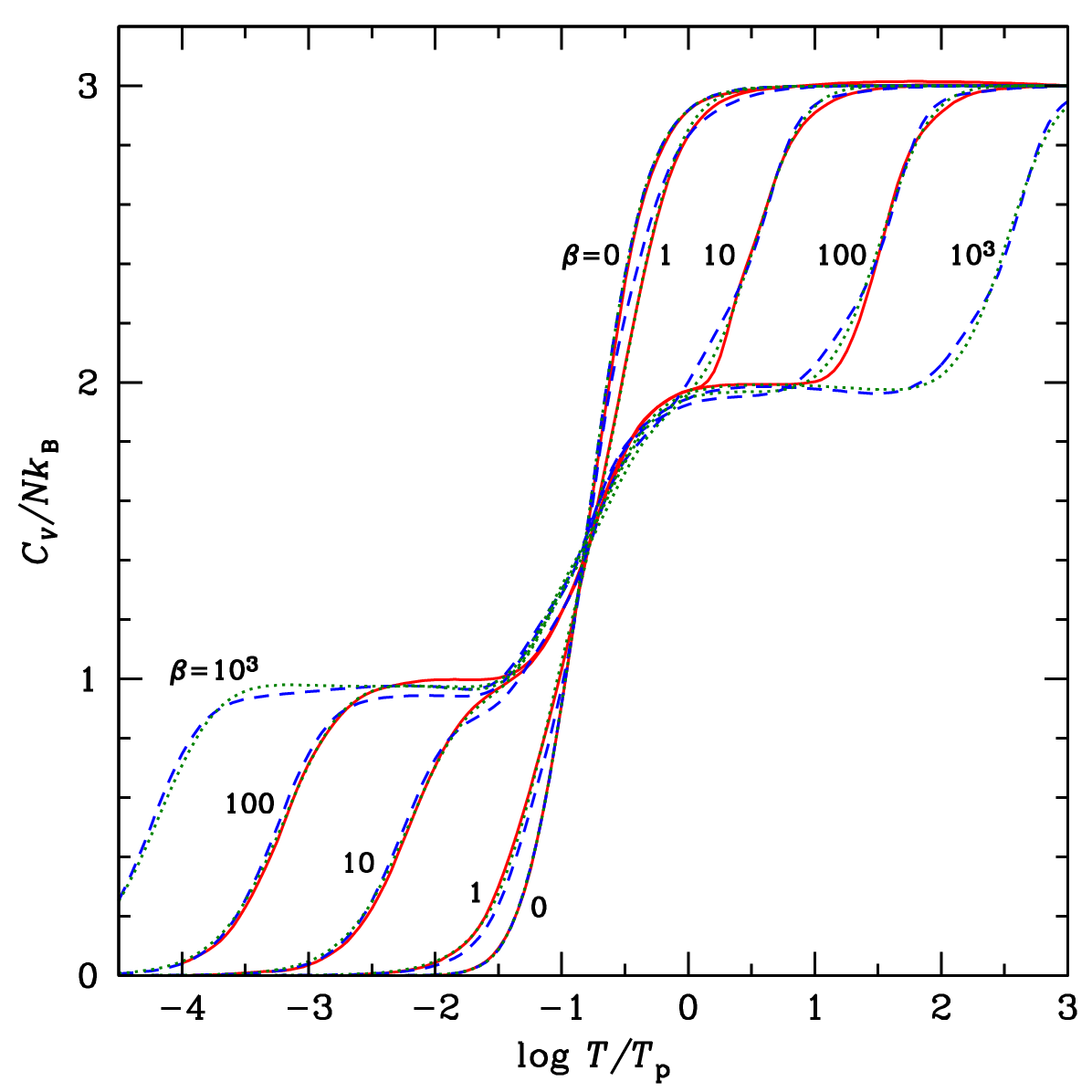}
\caption{
Thermal phonon contribution to the reduced heat capacity
$C_{V,\mathrm{lat}}/\Nion\kB T$ as a function of
$\log(T/\Tp)$ at $\beta=\hbar\omci/\kB\Tp = 0$, 1, 10, 100,
and $10^{3}$ (numbers near the lines). The analytical
approximation in \req{cfit1}) (short-dashed lines) is compared with
the numerical results of \citet{Baiko09} (solid lines for
$\beta=0$, 1, 10, and 100). The dotted lines correspond to
the first term on the r.h.s.\ of \req{cfit1}.
}
\label{fig:cth}
\end{figure}

The functions $\tilde{u}(\eta,\beta)$ and
$\tilde{s}(\eta,\beta)$ are displayed in Figs.~\ref{fig:uth}
and \ref{fig:sth}. Their accuracy is seen from a
comparison with the  numerical results \citep{Baiko09}, also
shown in the figures. However, if the complete consistency
of different thermodynamic functions is required,
Eqs.~(\ref{ufit}) and (\ref{sfit}) should not be used
directly, but should be first combined into 
$
   f_{\mathrm{th}}=\tilde{u}-\tilde{s}.
$
Then one can  calculate thermodynamic functions
by differentiating the function $f_{\mathrm{th}}(\eta,\beta)$. 
In this way we obtain, for example,
\bea&&
   u_{\mathrm{th}}=\frac{U_{\mathrm{th}} }{ \Nion\kB T } 
        = \tilde{u} + \Delta u,
\label{ufit1}
\\&&
  s_{\mathrm{th}}= {S_{\mathrm{th}} }/{\Nion\kB} =
     \tilde{s} + \Delta u,
\label{sfit1}
\\&&
\frac{C_{V,\mathrm{th}} }{\Nion\kB} =
   -
\left(\frac{\partial\tilde{s}}{\partial\ln\eta}\right)_\beta
   - \left(\frac{\partial\Delta u
         }{
           \partial\ln\eta}\right)_\beta,
\label{cfit1}
\\&&
\mbox{where~~}
\Delta u=
    \left(\frac{\partial\tilde{s}}{\partial\ln\eta}\right)_\beta
     - \left(\frac{\partial\tilde{u}}{\partial\ln\eta}\right)_\beta
        - \tilde{u}.
\eea
Note that the relation between the
phonon contributions to pressure and internal energy,
$2P_\mathrm{th}V=U_\mathrm{th}$, which is standard
for a nonmagnetized harmonic crystal, is invalid in the
strongly magnetized crystal because both dimensionless
arguments $\eta$ and $\beta$ depend on density.

Approximations in Eqs.~\ref{ufit1})\,--\,\ref{cfit1} are shown in
Figs.~\ref{fig:uth}\,--\,\ref{fig:cth}. Their reasonable
behavior beyond the range of available numerical data is
demonstrated by plotting them also at larger $\beta=10^3$
and $10^4$.

\subsection{Zero-point vibrations}

Because motions of the ions are confined by the magnetic
field in the transverse direction, they
exhibit quantum oscillations in the ground state
\citep{Landau30,LaLi-QM}. The energy of these
oscillations is
$\hbar\omci/2$ for every ion, which gives the term $\zeti/2$ in
\req{Fp}. In a Coulomb crystal, the motion of an ion
is confined in an effective potential well, centered
at its equilibrium lattice site. The total energy of the
zero-point quantum lattice oscillations $\Uq$
is given by \req{uq}.

In the case where the crystal is placed in a magnetic field,
$\Uq$ includes contributions due to both magnetic and
lattice confinements of the ion motion. However, since the
magnetic contribution $\Nion\hbar\omci/2$ is common in all
phase states, we  take it as the zero energy point in our
code and separate it from the lattice contribution that is
specific to the solid phase. Then \req{uq} becomes
\beq
   \Uq = \frac32\Nion\hbar\omp u'_1
         + \frac12 \Nion\hbar\omci,
\eeq
and in \req{flat} we have $1.5u_1\eta = 1.5u'_1\eta+\zeti/2$,
where we have defined
\beq
    u'_1 = u_1-\beta/3.
\eeq
The reduced frequency moment $u'_1$ still depends on
$\beta$, because the character of ion vibrations is affected
by the magnetic field (they become essentially
one-dimensional if $B$ is extremely large), but the latter
dependence is relatively weak. Having extracted $u'_1$ from
the available numerical results for $u_1$
\citep{Baiko09,BYak},
we can represent it by the simple interpolation
\beq
   u'_1(\beta) = \frac{u_1^0 + 1.27\beta^{9/8} u_1^{\infty}
           }{
             1+1.27\beta^{9/8}},
\eeq
where $u_1^0$ is the zero-field value and $u_1^{\infty}$ is the limit of $u'_1(\beta)$ at
$\beta\to\infty$. Only one of the three phonon branches
contributes to $u'_1$ in the latter limit, therefore
$u_1^{\infty}\sim u_1^0/3$. For the bcc crystal, 
$u_1^0=0.5113875$, whereas $u_1^{\infty}$ varies between
0.18 and 0.19 depending on the orientation of the lattice in
the magnetic field. In Fig.~\ref{fig:u1} we show $u'_1$ and
the logarithm (base 10) of $u_1$ versus $\beta$.

\begin{figure}
\includegraphics[width=\columnwidth]{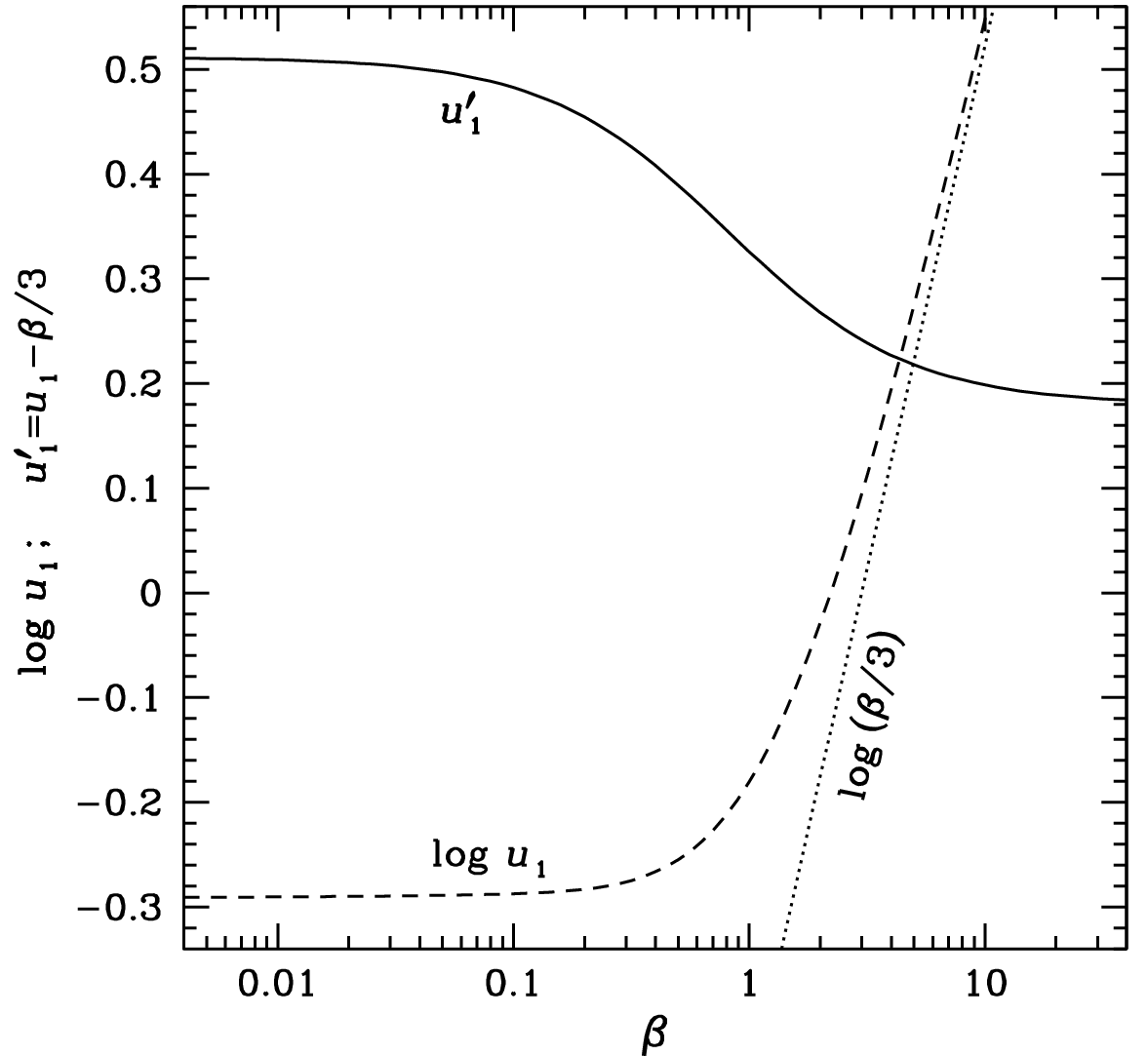}
\caption{
Reduced first moment of phonon frequencies $u'_1$ (solid
line) and $\log u_1$ (dashed line) as functions of
$\beta=\hbar\omci/\Tp$. The dotted line shows the asymptote
$\log(\beta/3)$.
}
\label{fig:u1}
\end{figure}

\citet{UsovGU} noticed that the energy of a crystal depends
on its orientation in a strong magnetic field. However,
numerical calculations \citep{Baiko-disser,Baiko09} show
that this dependence is very weak. For example, the
difference $\Delta u_1$ between the values of $u_1$ for two
orientations, where the field lines connect an ion with its
nearest neighbor in the first case and with a next-order
nearest neighbor in the second case, is approximately
\beq
   \Delta u_1 =
         \frac{\,\beta^{15/4}}{(1+\beta^{3/2})^{5/2}}\,
              (\Delta u_1)_\mathrm{max}
\eeq
with saturation level $(\Delta
u_1)_\mathrm{max}=0.0064$ for the bcc lattice.

\subsection{Comparison with the Baiko-Yakovlev fit}

After the present work was completed, we became aware
of an independent study by \citet{BYak}, who developed
another set of approximations for the free energy of the
harmonic Coulomb crystal in a magnetic field. They presented
the free energy as a sum of three terms corresponding to the
contributions from each of the three phonon modes in the bcc
crystal. Thus each of these terms has a clear physics
meaning, while our fitting expressions give only the total
contribution, which cannot be easily decomposed to three
parts corresponding to the separate phonon modes.

Unlike our fit, the fit of \citet{BYak} does not exactly
reproduce the very accurate results of \cite{Baiko-ea01} in
the limit $\beta\to0$. At finite $\beta$, both sets of
fitting expressions accurately reproduce the asymptotes at
$T\ll\Tp$ and $T\gg\Tp$ and have similar accuracies within
several percent points in the intermediate range
$0.1\Tp/\beta\lesssim T\lesssim10\Tp$.
Meanwhile, our approximation is simpler:
the Baiko-Yakovlev approximation contains 27
independent numerical fitting parameters, whereas our fits
\ref{ufit} and \ref{sfit} contain together only 9 such
parameters.

\section{Examples and discussion}
\label{sect:examples}

\subsection{Thermodynamic functions}

Characteristic features of the EOS can be seen in
Fig.~\ref{fig:eosmag}. Here, we have chosen the plasma
parameters that are typical for outer envelopes of isolated
neutron stars: we consider fully-ionized iron ($Z=26$,
$A=56$) at $T=10^7$~K and $B=10^{12}$~G 
(for illustration, the density range is extended to
$\rho\lesssim10^5$ neglecting the bound states
that can be important in this $\rho$\,--\,$T$ domain). 
We plot the normalized
pressure $p=P/\nion\kB T$, entropy $S/\Nion\kB$, heat
capacity $c_V=C_V/\Nion\kB$, and 
logarithmic derivatives of pressure
$\chi_\rho$ and $\chi_T$ as functions of density. Dashed
lines show these functions in the absence of quantizing
magnetic field. The vertical dotted lines marked by numbers
separate different characteristic domains, consecutively
entered with increasing density: onset of electron
degeneracy at $B=0$ ($\TF^\mathrm{(0)}=T$) and at
$B=10^{12}$~G ($\TF=T$),  population of excited Landau
levels ($\rho=\rho_B$), melting point with formation of a
classical Coulomb crystal ($\Tm=T$), and quantum effects in the
crystal ($\Tp=T$}).

At low densities, the ideal-gas values are approached:
$p=1+Z$, $\chi_\rho=\chi_T=1$,
$c_V=(3+3Z)/2$ at $B=0$, and $c_V=(3+Z)/2$ at
$B=10^{12}$~G. The latter difference is because
at $\rho < \rho_B$ the electron gas is effectively
one-dimensional due to the strong magnetic quantization.

With increasing density, the reduced pressure $p$ first
decreases below its ideal-gas value due to the Coulomb
nonideality and then increases due to the electron
degeneracy. The increase occurs earlier at zero field than
in the strong magnetic field, because of the delayed onset
of the degeneracy (Sec.~\ref{sect:mag-id}). When $\rho >
\rho_B\approx1.5\times10^4$ \gcc, the thermodynamic
functions approach their zero-field values. The gradually
decreasing oscillations correspond to consecutive filling of
the electron Landau levels. The magnetic field $B=10^{12}$~G
does not affect the ion contributions in
Fig.~\ref{fig:eosmag}, because it is nonquantizing for the
iron nuclei at $T=10^7$~K ($\zeti=0.00342$).

The liquid-solid phase transition occurs in
Fig.~\ref{fig:eosmag} at $\rho\approx8.25\times10^4$ \gcc,
where we adopt the classical OCP melting condition
$\Gamma=175.2$ (Paper~I). With further increase in
density ($\rho\gtrsim10^6$) the degeneracy becomes so
strong that the energy and pressure are nearly independent
of $T$ and $\chi_T$ strongly decreases. The normalized heat
capacity gradually tends to its value $c_V=3$ characteristic
of the classical simple crystal. At still higher density
the ion motions become quantized ($\Tp \gg T$) which leads
to the further decrease in the heat capacity and the entropy. 

\begin{figure}
\includegraphics[width=\columnwidth]{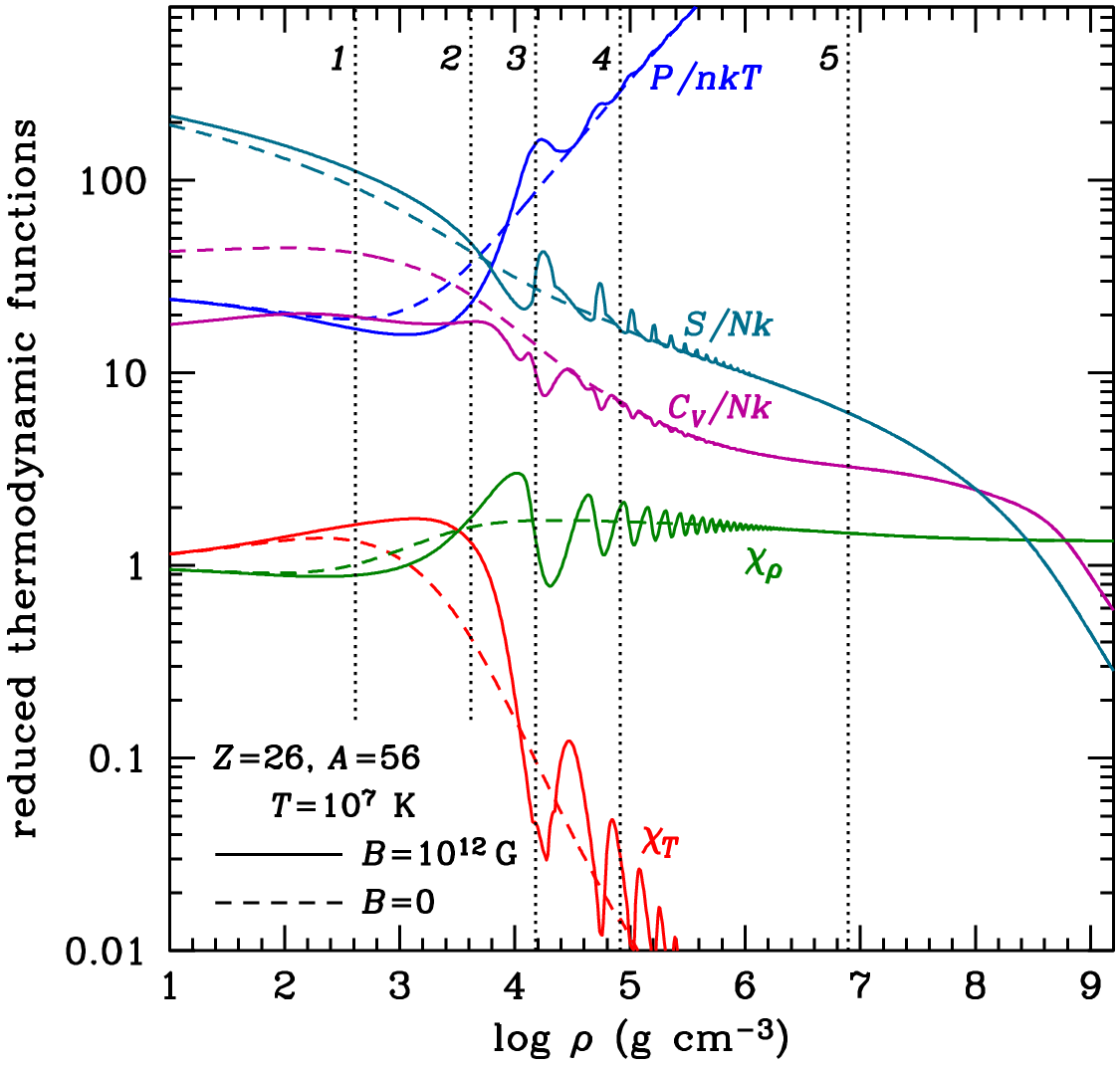}
\caption{
Reduced thermodynamic functions $P/\nion\kB T$,
$S/\Nion\kB$, $C_V/\Nion\kB$, $\chi_\rho$, and $\chi_T$ for
a fully-ionized nonmagnetic (dashed lines) and magnetized
($B=10^{12}$~G, solid lines) iron plasma at $T=10^7$~K.
The vertical dotted lines mark the densities at which (\emph{1})
$\TF^{(0)}=T$, (\emph{2}) $\TF=T$, (\emph{3})
$\rho=\rho_B$, (\emph{4}) $\Gamma=\Gammam$, and
(\emph{5}) $\Tp=T$.
}
\label{fig:eosmag}
\end{figure}

\begin{figure}
\includegraphics[width=\columnwidth]{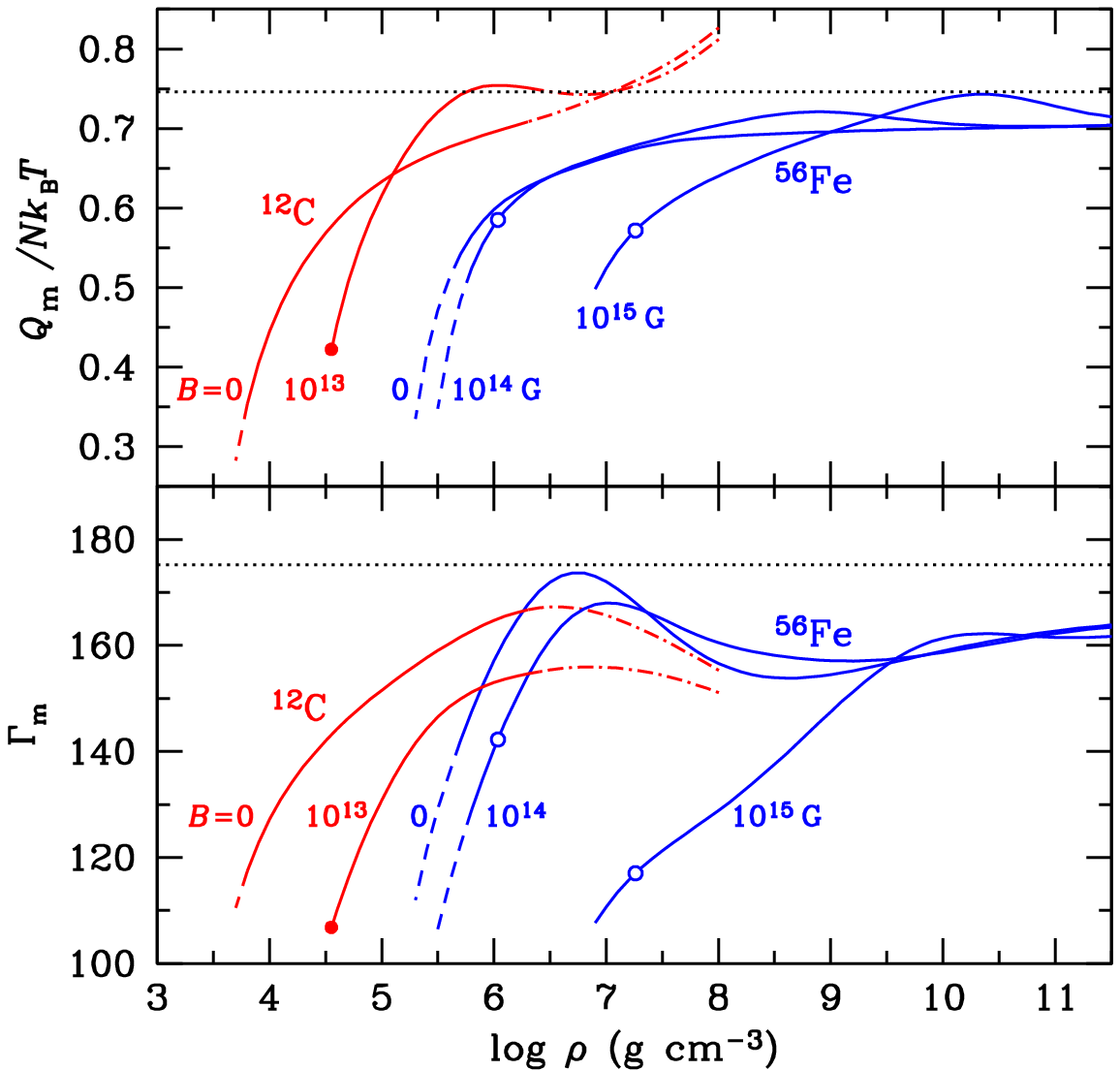}
\caption{Characteristics of the melting transition of
nonideal carbon and iron plasmas at different field
strengths $B$ (marked near the curves). Lower panel: the
value $\Gammam$ of the Coulomb coupling parameter $\Gamma$
at the melting point as function of mass density $\rho$.
Upper panel: normalized latent heat per ion at the melting
transition. The dot-dashed and dashed segments of the curves
correspond to the domains of nonperturbative quantum effects
($T < 0.5\,\Tp$) and electron response ($Z^2\,\textrm{Ry} >
0.1\, \EF$), respectively. The dotted horizontal lines mark
the OCP values. The filled and open circles mark the
positions of the real and virtual condensed surfaces (see
text in Sect.~\ref{sect:cond}).
}
\label{fig:melt}
\end{figure}

\subsection{Melting}

The electron polarization, ion quantum effects, and 
quantizing magnetic field can shift the melting temperature.
The lower panel of Fig.~\ref{fig:melt} shows the Coulomb
coupling parameter $\Gamma$ at the melting (that is, the
value $\Gammam$ at which the free energies of the two phases
are equal to each other); the upper panel displays the
difference between the internal energies in the liquid and
solid phases at the melting point (the latent heat
$Q_\mathrm{m}=U_\mathrm{liq}-U_\mathrm{sol}$), divided by
$\Nion\kB T$. We plot the data for fully ionized $^{12}$C
and $^{56}$Fe at $B=0$ and $10^{13}$~G for carbon, $B=0$,
$10^{14}$~G, and $10^{15}$~G for iron. The density range
shown in the figure is typical for the outer envelopes of a
neutron star and is also relevant for white dwarfs.

The position of the melting point is very sensitive to the
accuracy of the free energies of the Coulomb liquid and
crystal (see, e.g., Paper~I). The polarization and
quantum corrections to the classical OCP free energy are not
known sufficiently well for finding the position of the
melting point in the whole interval of densities shown in
Fig.~\ref{fig:melt}. The dot-dashed curves in this figure
correspond to the domain where $T < 0.5\,\Tp$.
Here, the perturbation theory for the quantum effects in the
liquid phase becomes progressively inaccurate. The dashed
curves correspond to the domain where the binding energy of
the hydrogen-like ion exceeds 10\% of the Fermi energy,
$Z^2\,\textrm{Ry} > 0.1\, \EF$, which corresponds to 
$\ael\gtrsim0.6a_0/Z$. Here, the perturbation treatment of
the electron polarization in the crystal starts to be
inaccurate. In addition, the position of the melting point
cannot be traced to $\Gamma\lesssim100$\,--\,120, because
the available results for the anharmonic corrections to the
free energy of a Coulomb crystal
(Appendix~\ref{sect:OCPsol}) are
accurate only at larger $\Gamma$.
Nevertheless, we can evaluate $\Gammam$ in a certain
interval of densities for each type of ions (the solid
segments of the curves in the figure).

The values of $Q_\mathrm{m}$ in Fig.~\ref{fig:melt} roughly
(within a factor of two) agree with the OCP value
$Q_\mathrm{m}^\mathrm{OCP}=0.746\Nion\kB T$ at
$\Gammam=175.2$ and with the values used in theoretical
models of white dwarf cooling (e.g., \citealp{Hansen04} and
references therein). Most of the neutron star cooling models
currently ignore the release of the latent heat at the
crystallization of the neutron star envelopes (e.g.,
\citealp{YakovlevPethick}, and references therein).

Figure~\ref{fig:melt} shows that the strong
magnetic fields tend to decrease $\Gammam$ and thus
stabilize the Coulomb crystal, in qualitative agreement with
previous conjectures
\citep{Ruderman71,KaplanGlasser,UsovGU,Lai01}. 
At densities
$\rho\sim10^7$\,--\,$10^8$ \gcc{} corresponding to the
``sensitivity strip'' in the neutron-star cooling theory
\citep{YakovlevPethick}, the stabilization proves to be
significant at the magnetar field strengths
$B=10^{14}$\,--\,$10^{15}$~G. 
The results for $^{56}$Fe in this $B$ interval, shown in the
lower panel of Fig.~\ref{fig:melt}, can be roughly (within
10\%) described by the formula
$\Gammam(B)\approx\Gammam(0)/(1+0.2\,\beta)$.
At the typical pulsar field
strengths, $B=10^{12}$\,--\,$10^{13}$~G, the effect is
noticeable at lower densities. However, these conclusions
remain preliminary in the absence of an evaluation of the
magnetic-field effects on the anharmonic corrections.
In view of the limited applicability and incompleteness of
the evaluation of $\Gammam$ with account of the
quantum, polarization, and magnetic effects, in applications
we use the classical OCP value $\Gammam=175.2$ as
the fiducial melting criterion.

\subsection{Magnetic condensation}
\label{sect:cond}

\citet{Ruderman71} suggested that the strong magnetic field
may
stabilize molecular chains (polymers) aligned with the
magnetic field and eventually
 turn the surface of a neutron star into the metallic
solid state. Later studies have provided support for this
conjecture, although the critical temperature
$T_\mathrm{crit}$, below which this condensation occurs,
remains 
very
 uncertain. Condensed surface density $\rhos$
is usually estimated as
\beq
    \rho_{\mathrm{s},\xi} = 
           561\,\xi\,AZ^{-0.6}\,B_{12}^{1.2}
          \textrm{~~~\gcc},
\label{rho_s} 
\eeq 
where $\xi\sim1$ is an unknown numerical factor, which
absorbs the theoretical uncertainty
\citep{Lai01,MedinLai06}. The value $\xi=1$ corresponds to
the EOS provided by the ion-sphere model \citep{Salpeter61},
which is close to the uniform model of
\citet{Fushiki-ea}.
For comparison, the results of the zero-temperature
Thomas-Fermi model for $^{56}$Fe at
$10^{10}\mbox{~G}\leqslant B\leqslant 10^{13}$~G
\citep{Rognvaldsson-ea} can be 
approximated  (within 4\%) by $\rho_{\mathrm{s},\xi}$
with $\xi\approx0.2+0.01/B_{12}^{0.56}$, whereas the
finite-temperature Thomas-Fermi model of
\citet{Thorolfsson-ea} does not predict magnetic
condensation at all.
Our EOS for partially ionized hydrogen plasmas in strong
magnetic fields \citep{PCS,PC04} exhibits a phase transition
with  $T_\mathrm{crit}\approx3\times10^5\,B_{12}^{0.39}$~K
and critical density $\rho_\mathrm{crit} \approx
143\,B_{12}^{1.18}\mbox{~\gcc} \sim \rho_\mathrm{s,0.25}$ at
$1\lesssim B_{12}\lesssim10^3$. According to another study
\citep{LaiSalpeter97,Lai01}, $T_\mathrm{crit}$ for hydrogen
is several times smaller.

\citet{MedinLai06} performed density-functional calculations
of the cohesive energy $Q_\mathrm{s}$ of the condensed phases of
H, He, C, and Fe in strong magnetic fields. A comparison
with previous density-functional calculations of other
authors prompts that $Q_\mathrm{s}$ may vary within a factor of
two at $B_{12}\gtrsim1$, depending on the approximations
(see \citealp{MedinLai06} for references and discussion). In
a subsequent study, \citet{MedinLai07} calculated the
equilibrium densities of saturated vapors of He, C, and Fe
atoms and polymers above the condensed surfaces, and 
obtained $T_\mathrm{crit}$ at several values of $B$ 
by
equating the vapor density to $\rhos$.
 Unlike
the previous authors, \citet{MedinLai06,MedinLai07} have
taken the electronic band structure of the condensed matter
into account self-consistently, but they did not allow for
the atomic motion across the magnetic field and mostly
neglected the contributions of the excited atomic and
molecular states in the gaseous phase.
\citeauthor{MedinLai07} obtained $\rhos$  assuming that the
linear molecular chains form a
rectangular array with sides $2R$ in the plane
 perpendicular to ${\bm B}$,
and that the distance $a$ between the nuclei along ${\bm B}$
in the condensed matter remains the same as in the
gaseous phase, so that $\rhos=\mion/4aR^2$
\citep{Medin12}. Using Tables~III\,--\,V of
\citet{MedinLai06} for $a$ and $R$, we can describe their
results for the surface  density of $^{12}$C at $1\leqslant
B_{12} \leqslant 1000$ and  $^{56}$Fe at $5\leqslant B_{12}
\leqslant 1000$ by \req{rho_s} with
$\xi=0.517+0.24/B_{12}^{1/5}\pm0.011$ and $\xi=0.55\pm0.11$,
respectively.

 \citet{MedinLai07}
found that the critical temperature is
$T_\mathrm{crit}\approx0.08 Q_\mathrm{s}/\kB$. 
Their numerical results
for He, C, and Fe can be roughly (within a factor of
1.5) described
as
$T_\mathrm{crit} \sim 5\times10^4\,Z^{1/4}\,B_{12}^{3/4}$~K
at $1\lesssim B_{12}\lesssim1000$. For comparison, the
results of \citet{LaiSalpeter97}
 for H at
$10\lesssim B_{12}\lesssim500$
 suggest $T_\mathrm{crit}
\sim 1.6\times10^4\,B_{12}^{0.7}$~K. The discrepancies between
different estimates 
of $\rhos$ and $T_\mathrm{crit}$
reflect the current theoretical
uncertainties.

The filled dots on the curves for magnetized carbon in both
panels of Fig.~\ref{fig:melt} correspond to the condensed
surface position in the fully-ionized plasma model.
In this model
$T_\mathrm{crit}\approx2.5\times10^5\,Z^{0.9}\,B_{12}^{0.4}$~K
and $\rho_\mathrm{crit} \approx \rho_\mathrm{s,0.47}$. With
decreasing temperature $T$ below $T_\mathrm{crit}$, the
surface density increases and tends to the limit 
$\rho_\mathrm{s,1}$ given using \req{rho_s} as 
$
   \rhos \approx \rho_\mathrm{s,1}
   / [1+1.1\,(T/T_\mathrm{crit})^5]
$
(cf.~Fig.~\ref{fig:diamag}).
At
smaller densities, there is a thermodynamically unstable
region in this model, therefore the curves 
in Fig.~\ref{fig:melt} are not continued
to the left beyond this point. For the magnetized iron
model, the melting curve does not cross the surface
because $\Tm>T_\mathrm{crit}$. In this case, the open
circles mark the density that the condensed surface would
have at much smaller temperatures $T\sim10^6\mbox{~K}\ll
T_\mathrm{crit}$. The parts of the curves to the left of the
open circles cannot be reached in a stationary stellar
envelope. 

\subsection{Thermal structure of a magnetar envelope}

\begin{figure}
\includegraphics[width=\columnwidth]{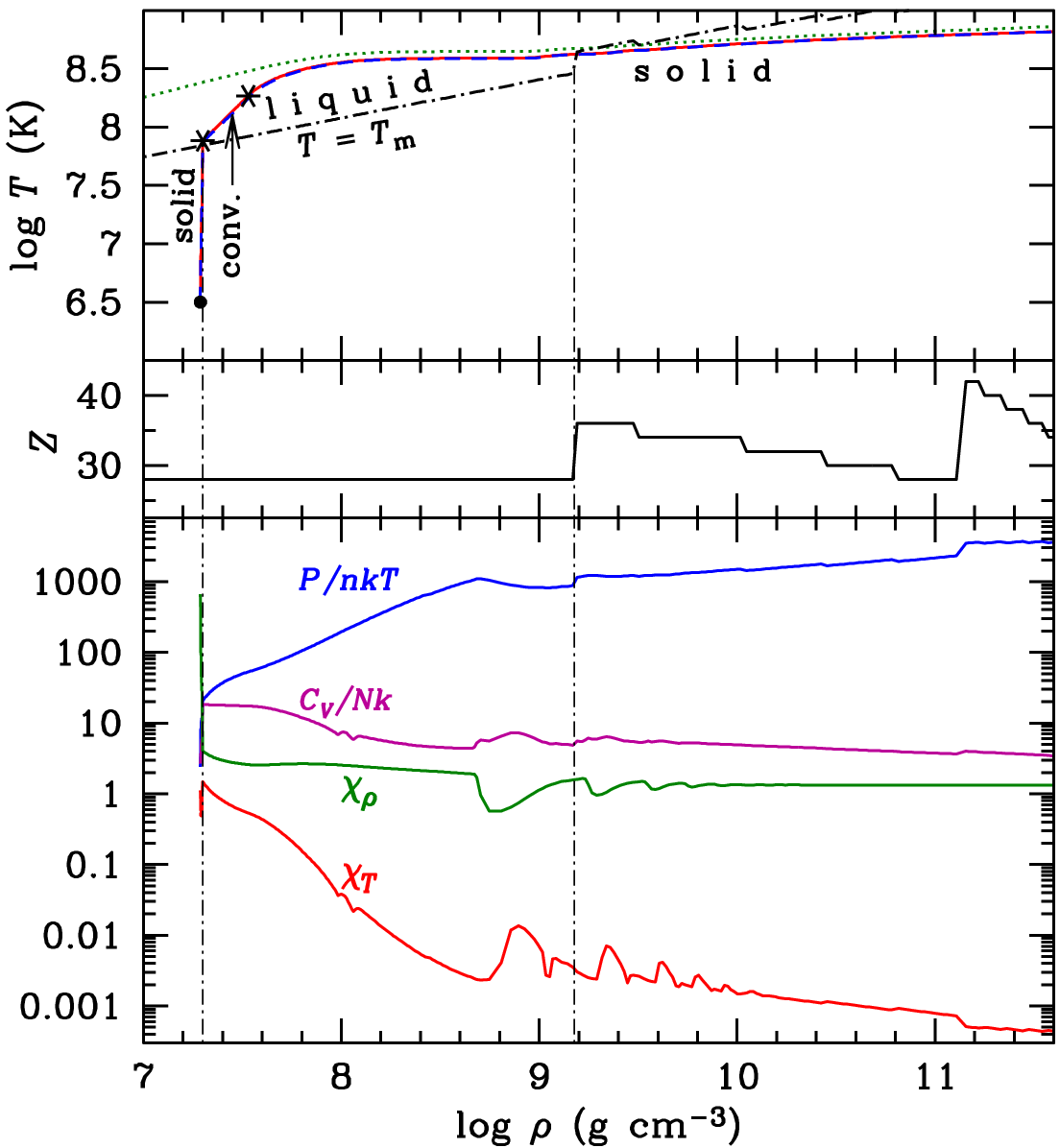}
\caption{Structure of a magnetar envelope having the
ground-state nuclear composition,
the effective temperature $10^{6.5}$~K,
and magnetic field $B=10^{15}$~G,
inclined at $45^\circ$ to the surface.
Top panel:
thermal profiles calculated using the present EOS (solid red
line) and the EOSs where the Coulomb nonideality is
either neglected (dotted green line) or treated without account of
magnetic quantization (dashed blue line, which is superposed on
the solid red line). The melting temperature
is drawn by the oblique dot-dashed line. Thin vertical
dot-dashed lines mark the points of phase transitions
from solid to liquid and back to solid state. Asterisks
mark the ends of the convective segment, which is
indicated by the arrow.
Middle panel:
ion charge \citep{Ruester}.
Bottom panel:
reduced pressure, heat
capacity, and logarithmic derivatives of pressure.
}
\label{fig:eospw}
\end{figure}

The results presented above have a direct application to the
calculations of the thermal and mechanical structure of
neutron-star envelopes with strong magnetic fields.
Figure~\ref{fig:eospw} illustrates the structure of a
typical magnetar envelope with the ground-state nuclear
composition. For illustration we have assumed that the
magnetar has mass $1.4\,M_\odot$ and radius 12 km, 
and the considered patch of the stellar surface has
effective
temperature $10^{6.5}$~K and magnetic
field $B=10^{15}$~G inclined at $45^\circ$.
The top panel shows the thermal structure of the envelope,
which has been calculated by numerical solution of the
system of heat balance equations, taking the general
relativity effects and neutrino emission into account
\citep{PCY}. The middle panel presents the ion charge $Z$ as
function of $\rho$ \citep{Ruester}. In the bottom panel
(analogous to Fig.~\ref{fig:eosmag}) we plot several reduced
thermodynamic functions of $\rho$ and $T$ along the thermal
profile (i.e., taking $T$ from the top
panel), starting at the condensed solid surface.

The temperature quickly grows 
at the solid surface
and reaches the melting point at
the depth
$z\approx7$~cm. Thus, at the given conditions, the liquid ocean of
a magnetar turns out to be covered by a thin layer of
``ice'' (solid substance). 
We treat the solid crust as immobile, but the liquid layer
below the ``ice'' is convective up to the depth $z\sim1$~m. 
We treat the convective heat transport through this layer in
the adiabatic approximation \citep{Schwarzschild}. The
change of the heat-transport mechanism from conduction to
convection  causes the break of the temperature profile at
the melting point. We underline that this treatment is
only an approximation. In reality, the superadiabatic growth
of temperature can lead to a hydrostatic instability of the
shell of ``ice'' and eventually to its cracking and
fragmentation into turning-up ``ice floes''. This 
can result in transient enhancements of the thermal
luminosity of magnetars.

The temperature profile flattens
with density increase, and the Coulomb plasma freezes again
at the interface between the layers of $^{66}$Ni and
$^{86}$Kr at $\rho=1.5\times10^9$ \gcc{} ($z=73.8$~m). These
phase transitions do not cause any substantial breaks in 
$\chi_\rho$, $\chi_T$, or $C_V/\Nion\kB$, 
because the Coulomb plasmas have
similar structure factors in the liquid and crystalline
phases in the melting region \citep[cf.][]{Baiko-ea98}. 

At the boundaries between layers composed of different
chemical elements, the reduced thermodynamic functions do
not exhibit substantial discontinuities, except for the
abrupt increases in $P/\nion\kB T$ at the interfaces
$^{66}$Ni/$^{86}$Kr ($\rho=1.5\times10^9$ \gcc) and
$^{78}$Ni/$^{124}$Mo ($\rho=1.32\times10^{11}$ \gcc), which
are caused by the decreases in $\nion$ with the large jumps
in $A$ (of course, the non-normalized pressure $P$ is
continuous). The specific heat per ion $C_V/\Nion\kB$ is
almost continuous at these interfaces, which means that heat
capacity of unit volume abruptly decreases. The drop in
$\chi_T$ at the $^{78}$Ni/$^{128}$Mo interface is due to the
same decrease in $\nion$, which leads to the decrease in the
ionic contribution that mostly determines $\partial
P/\partial T$ at the strong degeneracy.

The oscillations of the reduced thermodynamic functions
(most noticeable for $\chi_\rho$ and $\chi_T$) correspond to
consecutive population of excited Landau levels by
degenerate electrons with density increase, analogous to the
oscillations in Fig.~\ref{fig:eosmag}.

The magnetic effects on the nonideal part of the plasma
thermodynamic functions have almost no influence on the
temperature profile in the magnetar envelope, as illustrated
in the upper panel of Fig.~\ref{fig:eospw} where the
corresponding solid and dashed lines virtually coincide. For
comparison, the dotted line in the upper panel shows the result of a
calculation totally neglecting the Coulomb nonideality. In
this case, the profile is quite different at low densities,
where there is no longer a solid surface. However, even in
this case the thermal profile is almost the same at large
$\rho$. This means that the Coulomb nonideality has a minor
impact on the relation between the internal and
effective temperatures and therefore on the cooling curves
\citep{YakovlevPethick}, but it can be important for the
shape of the thermal spectrum (cf., e.g.,
\citealp{Potekhin-ea12}).

\section{Conclusions}
\label{sect:concl}

We have systematically reviewed analytical approximations for
the EOS of fully-ionized electron-ion plasmas in magnetic
fields and described several improvements to the previously
published approximations, taking nonideality
attributable to ion-ion, electron-electron, and electron-ion
interactions into account. The presented formulae are applicable in a
wide range of plasma parameters, including the domains of
nondegenerate and degenerate, nonrelativistic and
relativistic electrons, weakly and strongly coupled Coulomb
liquids, classical and quantum Coulomb crystals. As an
application, we have calculated and discussed the behavior
of thermodynamic functions, melting, 
and latent heat at crystallization of strongly coupled
Coulomb plasmas with the parameters appropriate for cooling
white dwarfs and envelopes of nonmagnetized and strongly
magnetized neutron stars. We have also shown that a typical
outer envelope of a magnetar can have a liquid layer beneath
the solid surface.

The Fortran code that realizes the analytical approximations
described in this paper is available at
\texttt{http://www.ioffe.ru/astro/EIP/} and at the CDS via
anonymous ftp to cdsarc.u-strasbg.fr (130.79.128.5) or via
\texttt{http://cdsweb.u-strasbg.fr/cgi-bin/qcat?J/A+A/}.

\begin{acknowledgements}
We are grateful 
to Jos\'e Pons for pointing out a bug in a previous version
of one of our subroutines,
to D.~A.~Baiko and D.~G.~Yakovlev for making
their results available to us prior to publication
and for useful discussions,
and to D.~G.~Yakovlev for valuable
remarks on a preliminary version of this paper.
The work of A.Y.P.\ was partially supported 
by the
Ministry of Education and Science of the Russian Federation
(Agreement No.\,8409, 2012),
the Russian Foundation for Basic Research (RFBR grant
11-02-00253-a), and the Russian Leading Scientific
Schools program (grant NSh-4035.2012.2). 
\end{acknowledgements}

\appendix

\section{Nonideal part of the free energy
of electrons}
\label{sect:Fee}

\citet{Tanaka85} calculated the interaction energy
of the electron fluid at finite $T$ and
presented a fitting formula that reproduced their numerical
results as well as the results of other authors in various limits.
Subsequently the behavior of the fit at $T\ll\TF$ was improved
by \citet{IIT}. The result reads
\bea&&\hspace*{-1em}
   \frac{F_{ee} }{ \kB T \Nel } = - g \Gamme - \frac{2A}{f}\,\sqrt\Gamme
\nonumber\\&&
   +
       \frac{2 (dB+CA)}{ f D }
  \left[\arctan\!\left(\! \frac{2f\,\sqrt\Gamme + d }{ D} \!\right)\!
      - \arctan\!\left(\frac{d}{D}\!\right) \right]
\nonumber\\&&
      - \left( \frac{B}{ f} - \frac{dA }{ f^2} \right)
       \ln ( f\Gamme+d\sqrt\Gamme+1),
\label{IIT}
\eea
where $A=b-gd$, $B=a-g$,
$C=2-d^2/f$, $D=\sqrt{4f-d^2}$, 
and $a$, $b$, $d$, $f$, and
$g$ are the following functions of $\theta=T/\TF$ (at $B=0$):
\bea
    a &=& \left( \frac{9}{4\pi^2} \right)^{1/3}
       \tanh\frac1\theta
\nonumber\\&&\times
      \frac{ 0.75 + 3.04363 \theta^2 -0.09227\theta^3 + 1.7035\theta^4
         }{ 1+8.31051\theta^2 +5.1105\theta^4}\,,
\nonumber\\
    b &=& 
        \frac{ 0.341308 + 12.0708\,\theta^2 + 1.148889 \,\theta^4
     }{ 1 + 10.495346 \,\theta^2 + 1.326623 \,\theta^4 }
      \sqrt\theta\,\tanh\frac{1}{\sqrt\theta}\,,
\nonumber\\
    d &=& \!\!
        \frac{0.614925 + 16.996055 \theta^2 + 1.489056 \theta^4
      }{ 1 + 10.10935 \,\theta^2 + 1.22184 \,\theta^4 }
      \sqrt\theta\,\tanh\frac{1}{\sqrt\theta}\,,
\nonumber\\
    f &=& 
        \frac{0.539409 + 2.522206 \,\theta^2 + 0.178484 \,\theta^4
      }{ 1 + 2.555501 \,\theta^2 + 0.146319 \,\theta^4 }
      \,\theta\,\tanh\frac{1}{\,\theta}\,,
\nonumber\\
    g &=& 0.872496 + 0.025248\,\exp(-1/\theta).
\nonumber
\eea
The accuracy of \req{IIT} is 1\%.

In a quantizing magnetic field, we replace the argument
$\theta$ in these expressions by the quantity $\theta^\ast$
defined by \req{theta-ast}, as explained in
Sect.~\ref{sect:ee-mag}.

\section{Nonideal part of the free energy
of ions in the rigid background}

\subsection{Coulomb liquid}
\label{sect:OCPliq}

For the reduced free energy
$f_\mathrm{ii}\equiv {F_\mathrm{ii}}/{\Nion\kB T}$ of the
classic OCP, we have the following analytical formula
(Paper~I):
\bea
  f_\mathrm{ii}^{(0)} 
   &= &
   A_1\Big[\sqrt{\Gamma(A_2+\Gamma)}
           - A_2 \ln\!\left(\sqrt{\Gamma/ A_2}
           +\sqrt{1+\Gamma/ A_2}\right)\Big]
 \nonumber\\&&
          + 2A_3\left[\sqrt{\Gamma}
           -\arctan\sqrt{\Gamma} \right]
\nonumber\\
         &+& \!\!\! B_1\left[\Gamma 
         - B_2\ln\left(1+\frac{\Gamma}{ B_2}\right)\right]
 +\frac{B_3}{2}\,\ln\left(1+\frac{\Gamma^2}{ B_4}\right),
\label{fition}
\eea
where $A_1=-0.907347$, $A_2=0.62849$, $B_1=0.0045$, $B_2=170$,
$B_3=-8.4\times10^{-5}$, $B_4=0.0037$, and
$A_3=-\sqrt{3}/2-A_1/\sqrt{A_2}$.
The derivative
\beq
      \frac{\partial f_\mathrm{ii}^{(0)}}{\partial\ln\Gamma}
         = \Gamma^{3/2}
   \left[\frac{A_1}{\sqrt{\Gamma+A_2}} 
   + \frac{A_3}{\Gamma+1}\right]
      + \frac{B_1\,\Gamma^2 }{\Gamma+B_2}
      + \frac{B_3\,\Gamma^2 }{\Gamma^2+B_4},
\eeq
reproduces Monte Carlo
calculations of the reduced internal energy
${U_\mathrm{ii}}/{\Nion \kB T}$
at $1\leq\Gamma\leq190$ \citep{Caillol} within the accuracy
of these calculations, $\lesssim10^{-3}$. For any values of
the coupling parameter in a liquid OCP,
$0\leqslant\Gamma\lesssim200$, the 
fractional error of the approximation (\ref{fition})
does not exceed $2\times10^{-4}$.

The classical treatment of ion motion is justified at
$T\gg\Tp$. One can extend the applicability range of the
analytical EOS to $T\sim\Tp$ by
the Wigner-Kirkwood quantum corrections
\citep{Wigner32,LaLi-SP1}.
The lowest-order correction to the reduced free energy
is
\beq
   f_\mathrm{ii}^{(2)}=\eta^2/24.
\label{fq}
\eeq
The next-order correction
$\propto\hbar^4$ was obtained by \citet{HV75}.
It can be written as
\bea
   f_\mathrm{ii}^{(4)} &\approx&
    \bigg(\!-2.085\times10^{-4} - \frac{2.411\times10^{-4}}{\Gamma^{1/2}}
    - \frac{0.001288}{\Gamma}
\nonumber\\&&
    +\frac{1.353\times10^{-4}}{\Gamma^{3/2}}
    - \frac{0.002476}{\Gamma^2}-\frac{0.00276}{\Gamma^{5/2}}\bigg)
    \,\eta^4.
\label{WK4}
\eea
This expression, unlike \req{fq}, is not exact.
Both corrections have
limited applicability, because as soon as $\eta$ becomes large,
the Wigner expansion diverges and
the plasma forms a quantum liquid, whose free energy
is not known in an analytical form. Therefore we use
only the lowest-order correction (\ref{fq}),
 i.e., $f_\mathrm{ii}\approx
f_\mathrm{ii}^{(0)}+f_\mathrm{ii}^{(2)}$.
In a magnetic field, \req{fq} is replaced by \req{fq-mag}
(Sect.~\ref{sect:fq-mag}).

\subsection{Coulomb crystal}
\label{sect:OCPsol}

The reduced free energy of an OCP in the crystalline phase
is given by \req{flat}, where the first three terms describe
the harmonic lattice model \citep{Baiko-ea01}. For the bcc
crystal, we have $C_0=-0.895\,929\,255\,68$ and
$u_1=0.511\,3875$, and for $f_\mathrm{th}$ the following
fitting formula can be used:
\beq
   f_\mathrm{th} = \sum_{k=1}^3\ln\left(1-e^{-\alpha_k\eta}\right) -
  \frac{A(\eta)}{ B(\eta)},
\label{HLfit}
\eeq
where $\alpha_1=0.932446$,
$\alpha_2=0.334547$, $\alpha_3=0.265764$,
\bea
 A(\eta) &=& 1 + 0.1839\,\eta+ 0.593\,586\,\eta^2+0.005\,4814\,\eta^3
\nonumber\\&&
   +5.018\,13\times10^{-4}\,\eta^4
   +3.9247\times10^{-7}\,\eta^6
\nonumber\\&&
   +5.8356\times10^{-11}\,\eta^8,
\nonumber
\\
B(\eta) &=& 261.66+7.079\,97\,\eta^{2}+0.0409\,484\,\eta^{4}
\nonumber\\&&
+ 3.973\,55\times10^{-4}\,\eta^{5}
+5.111\,48\times10^{-5}\,\eta^{6}
\nonumber\\&&
     +  2.197\,49\times10^{-6}\,\eta^{7}
     +1.866\,985\times10^{-9}\,\eta^{9}
\nonumber\\&&
     +2.787\,72\times10^{-13}\,\eta^{11} .
\nonumber
\eea
The Taylor expansion of \req{HLfit} at small $\eta$ is
consistent with the Wigner correction \ref{fq}. However,
the next Taylor term $\sim\eta^3$ is absent in the Wigner
expansion, and therefore \req{HLfit} does not reproduce
higher-order Wigner corrections. Nevertheless, approximation
\ref{HLfit} is very accurate: it reproduces
the numerical results in \citet{Baiko-ea01} with fractional
deviations within $5\times10^{-6}$, and its first and second
derivatives reproduce the calculated contributions to the
internal energy and heat capacity with deviations up to
several parts in $10^5$. Other types of simple lattices are
described by the same expressions with slightly different
parameters \citep[see][]{Baiko-ea01}. 

Anharmonic corrections for Coulomb lattices were
studied in a number of works (see Papers~I and II for references).
In the classical regime $\eta\to0$, we have chosen one of
the 11
parametrizations proposed by \citet{FaroukiHamaguchi}:
\beq
 f_\mathrm{ah}^{(0)}(\Gamma) = -\sum_{k=1}^3
        \frac{a_k}{k\Gamma^k},
\label{Farouki}
\eeq
where  $a_1=10.9$, $a_2=247$, and $a_3=1.765\times10^5$.
A continuation to arbitrary $\eta$, which is consistent with
available analytical and numerical results for quantum
crystals, reads (Paper~II)
\beq
      f_\mathrm{ah} = 
          f_\mathrm{ah}^{(0)}(\Gamma)\,e^{-0.0112\eta^2}
             -0.12\,\eta^2/\Gamma\,.
\label{f_ah_fit}
\eeq

Superstrong magnetic fields can significantly change these
expressions under the conditions $\zeti\gtrsim1$ and
$\eta\gtrsim1$. Analytical approximations for the free
energy of a harmonic Coulomb crystal in quantizing magnetic
fields are derived in Sect.~\ref{sect:mag-sol}. Analogous
results for the anharmonic corrections are currently
unavailable.

\section{Electron polarization corrections}
\label{sect:ei2}

\subsection{Coulomb liquid}
\label{sect:ei-liq}

The screening contribution to the reduced free energy of the
Coulomb liquid at $0<\Gamma\lesssim300$ has been calculated
by the HNC technique and fitted by the expression
(Paper~I)
\beq
  f_\mathrm{ie} \equiv \frac{F_\mathrm{ie} }{ \Nion \kB T}
   = -\Gamme \,
   \frac{ c_\mathrm{DH} \sqrt{\Gamme}+
    c_\mathrm{TF} a \Gamme^\nu g_1 h_1
    }{
     1+\left[ b\,\sqrt{\Gamme}+ a g_2 \Gamme^\nu/\rs \right]
    \gr^{-1} },
\label{fitscr}
\eeq
where 
$
   c_\mathrm{DH} = (Z/\sqrt{3})
    \left[(1+Z)^{3/2}-1-Z^{3/2}\right]
$
ensures exact transition to the Debye-H\"uckel limit at $\Gamma\to0$,
$c_\mathrm{TF} = 
   (18/175)\,(12/\pi)^{2/3} 
   Z^{7/3}\left(1-Z^{-1/3}+0.2\,Z^{-1/2}\right)
$
fits the numerical data at large $\Gamma$ and
reproduces the Thomas-Fermi limit \citep{Salpeter61} 
at $Z\to\infty$,
the parameters
$   a = 1.11 \, Z^{0.475}$,
$
   b = 0.2+0.078 \,(\ln Z)^2,
$ and
$
   \nu = 1.16 + 0.08 \ln Z
$
provide a low-order approximation to $F_{ie}$ for intermediate
$\rs$ and $\Gamma$. The functions
\bea
   g_1 &=& 1 + 0.78 \left[
           21+ \Gamme(Z/\rs)^3 \right]^{-1} (\Gamme/Z)^{1/2},
\nonumber\\
   g_2 &=& 1+\frac{ Z-1 }{ 9}
         \left(1+\frac{1}{0.001\, Z^2+2\Gamme}\right)
         \frac{\rs^3 }{ 1+ 6 \, \rs^{2} }\,,
\nonumber
\eea
improve the fit at relatively large $\rs$.
Finally, the function
\[
 h_1 = \frac{1+ \xr^2 / 5 }{ 1+0.18\,Z^{-1/4} \xr
 + 0.37\, Z^{-1/2} \xr^2 +\xr^2 /5}
\]
is the relativistic correction, as is $\gr^{-1}$
in the denominator.

\subsection{Coulomb crystal}
\label{sect:ei-sol}

The screening contribution to the reduced free energy of the
Coulomb crystals was evaluated using the semiclassical
perturbation approach with an effective structure factor
(Paper~I) and fitted by the expression (Paper~II)
\beq
   f_\mathrm{ie} = -f_\infty(\xr)\,\Gamma
         \left\{ 1 + A(\xr)\,\big[Q(\eta)/\Gamma\big]^s
\right\},
\label{fitscr-sol}
\eeq
where
\bea
   f_\infty(x) &=& a_\mathrm{TF} Z^{2/3} b_1\,\sqrt{1+b_2/x^2},
\nonumber\\
   A(x) &=& \frac{ b_3+17.9 x^2 }{ 1+b_4 x^2 },
\nonumber\\
   Q(\eta) &=& \left[ { \ln\left(1+e^{(0.205\eta)^2} \right)
          }\right]^{1/2}\,\left[{
             \ln\left(e-(e-2)e^{-(0.205\eta)^2} \right)}
             \right]^{-1/2},
\nonumber
\eea
the parameter $a_\mathrm{TF} = 0.00352$ is related to
$c_\mathrm{TF}$ in \req{fitscr},
and parameters $s$ and $b_1$\,--\,$b_4$ depend on $Z$:
\bea
   s &=& \left[ 1+0.01\,(\ln Z)^{3/2} + 0.097\,Z^{-2} \right]^{-1},
\nonumber\\
  b_1 &=& 1 - 1.1866 \,Z^{-0.267} + 0.27\,Z^{-1},
\nonumber\\
  b_2 &=& 1 + \frac{2.25}{ Z^{1/3}}\,
      \frac{1+0.684\,Z^5+0.222\,Z^6 }{ 1+0.222\,Z^6},
\nonumber\\
  b_3 &=& 41.5/(1+\ln Z),
\nonumber\\
  b_4 &=& 0.395 \ln Z + 0.347\, Z^{-3/2}.
\nonumber
\eea
Here, the numerical parameters are given for the bcc
crystal;
their values for the fcc lattice are slightly different
(Paper~I).



\begin{thebibliography}{99}

\bibitem[Alastuey \& Jancovici(1980)]{AlaJanco}
Alastuey, A., \& Jancovici, B.
1980,
Physica A, 102, 327

\bibitem[Baiko(2000)]{Baiko-disser}
Baiko, D.~A. 
2000,
PhD thesis (St.\ Petersburg: Ioffe Phys.-Tech.\ Inst.) [in Russian]

\bibitem[Baiko(2002)]{Baiko02}
Baiko, D.~A. 
2002,
\pre, 66, 056405

\bibitem[Baiko(2009)]{Baiko09}
Baiko, D.~A.
2009,
\pre, 80, 046405

\bibitem[Baiko \& Yakovlev(2012)]{BYak}
Baiko, D.~A., \& Yakovlev, D.~G.
2012,
private communication

\bibitem[Baiko et al.(1998)]{Baiko-ea98}
Baiko, D.~A., Kaminker, A.~D., Potekhin, A.~Y., \& Yakovlev, D.~G.
1998,
\prl, 81, 5556

\bibitem[Baiko et al.(2001)Baiko, Potekhin, \& Yakovlev]{Baiko-ea01}
Baiko, D.~A., Potekhin, A.~Y., \& Yakovlev, D.~G.
2001,
\pre, 64, 057402

\bibitem[Banerjee et al.(1974)Banerjee, Constantinescu, \& Rehak]{BanerjeeCR}
Banerjee, B., Constantinescu, D.~H., \& Rehak, P.
1974,
\prd, 10, 2384

\bibitem[Berestetski{\u\i} et al.(1982)]{LaLi-QED}
Berestetski{\u\i}, V.~B., Lifshitz, E.~M., \&
Pitaevski{\u\i}, L.~P.,
1982,
Quantum Electrodynamics
(Oxford: Butterworth-Heinemann)

\bibitem[Blandford \& Hernquist(1982)]{BlandfordHernquist}
Blandford, R.~D., \& Hernquist, L.
1982,
J.\ Phys.\ C: Solid State Phys., 15, 6233

\bibitem[Blinnikov et al.(1996)Blinnikov, Dunina-Barkovskaya, \& Nadyozhin]{Blin}
Blinnikov, S.~I., Dunina-Barkovskaya,  N.~V. Nadyozhin, D.~K.
1996,
\apjs, {106} 171;
erratum: 1998, \apjs, 118, 603

\bibitem[Brami et al.(1979)Brami, Hansen, \& Joly]{BHJ79}
Brami, B., Hansen, J.~P., \& Joly, F.
1979,
Physica, 95A, 505

\bibitem[Broderick et al.(2000)Broderick, Prakash, and Lattimer]{Broderick}
Broderick, A., Prakash, M. \& Lattimer, J.~M.
2000,
\apj, 537, 351

\bibitem[Caillol(1999)]{Caillol}
Caillol, J.~M.
1999,
J.\ Chem.\ Phys., 111, 6538

\bibitem[Canuto \& Chiu(1968)]{CanutoChiu}
Canuto, V., \& Chiu, H.-Y.
1968,
Phys.\ Rev., 173, 1220

\bibitem[Chabrier \& Ashcroft(1990)]{ChabAsh}
Chabrier G., \& Ashcroft, N.~W.
1990,
\pra, 42, 2284

\bibitem[Chabrier \& Baraffe(2000)]{ChaBar}
Chabrier G., \& Baraffe, I.
2000,
\araa, 38, 337

\bibitem[Chabrier \& Potekhin(1998)]{CP98}
Chabrier, G., \& Potekhin, A.~Y.
1998,
\pre, 58, 4941

\bibitem[Chabrier et al.(2002)Chabrier, Douchin, \& Potekhin]{CDP02}
Chabrier, G., Douchin, F., \& Potekhin, A.~Y.
2002,
J.~Phys.: Condensed Matter, 14, 9133

\bibitem[Chandrasekhar(1957)]{Chandra}
Chandrasekhar, S.
1957,
{An Introduction to the Study of Stellar
Structure} 
(New York: Dover, 1957), Chap.~X.

\bibitem[Cornu(1998)]{Cornu}
Cornu, F.
1998,
\pre, 58, 5293

\bibitem[Danz \& Glasser(1971)]{DanzGlasser}
Danz, R.~W., \& Glasser, M.~L.
1971,
\prb, 4, 94

\bibitem[DeWitt et al.(1996)DeWitt, Slattery, \& Chabrier]{DWSC96}
DeWitt, H.~E., Slattery, W., \& Chabrier, G.
1996,
Physica A, 228, 21

\bibitem[DeWitt \& Slattery(2003)]{DWS03}
DeWitt, H.~E., \& Slattery, W.
2003,
Contrib.\ Plasma Phys., {43}, 279

\bibitem[Farouki \& Hamaguchi(1993)]{FaroukiHamaguchi}
Farouki, R.~T., \& Hamaguchi, S.
1993,
\pre, 47, 4330

\bibitem[Fortov(2009)]{Fortov}
Fortov, V.~E.
2009,
Phys. Usp., 52, 615

\bibitem[Fushiki et al.(1989)Fushiki, Gudmundsson, \& Pethick]{Fushiki-ea}
Fushiki, I., Gudmundsson,  E.~H., \& Pethick, C.~J.
1989,
\apj, 342, 958

\bibitem[Galam \& Hansen(1976)]{GalamHansen}
Galam S., \& Hansen, J.~P.
1976,
\pra, 14, 816

\bibitem[Girifalco(1973)]{Girifalco}
Girifalco, L.~A.
1973,
{Statistical Physics of Materials} 
(New York: Wiley-Interscience, 1973), App.~V.

\bibitem[Griffith(1999)]{Griffith}
Griffith, D.~J.
1999,
{Introduction to Electrodynamics}, 3rd ed.
(London: Prentice-Hall), Chap.~6.

\bibitem[Haensel et al.(2007)Haensel, Potekhin, \& Yakovlev]{NSB1}
 Haensel, P., Potekhin, A.~Y., \& Yakovlev, D.~G.
 2007,
Neutron Stars 1: Equation of State and Structure
(New York: Springer)

\bibitem[Hamaguchi et al.(1997)Hamaguchi, Farouki, \& Dubin]{Hama-Yukawa}
Hamaguchi, S., Farouki,  R.~T., \& Dubin,  D.~H.~E.
1997,
\pre, 56, 4671

\bibitem[Hansen(2004)]{Hansen04}
Hansen, B.~M.~S.
2004,
Phys.\ Rep., 399, 1

\bibitem[Hansen et al.(1977)Hansen, Torrie, \& Vieillefosse]{HTV77}
Hansen, J.~P., Torrie, G.~M., \& Vieillefosse, P.
1977,
\pra, {16}, 2153

\bibitem[Hansen \& Vieillefosse(1975)]{HV75}
Hansen, J.~P., \& Vieillefosse, P.
1975,
Phys.\ Lett. A, 53, 187

\bibitem[Hughto et al.(2012)]{Hughto-ea12}
Hughto, J., Horowitz, C.~J., Schneider, A.~S., et al.
2012,
\pre, accepted (arXiv:1211.0891)

\bibitem[Ichimaru et al.(1987)Ichimaru, Iyetomi, \& Tanaka]{IIT}
Ichimaru, S., Iyetomi, H., \& Tanaka, S.
1987,
Phys.\ Rep., {149}, 91

\bibitem[Jancovici(1962)]{Jancovici}
Jancovici, B.
1962,
Nuovo Cimento, 25, 428

\bibitem[Johnson \& Lippmann(1949)]{JohnsonLippmann}
Johnson, M.~H., \& Lippmann, B.~A.
1949,
Phys.~Rev., 76, 828

\bibitem[Kaplan \& Glasser(1972)]{KaplanGlasser}
Kaplan, J.~L., \& Glasser, M.~L.
1972,
\prl, 28, 1077

\bibitem[Kittel(1963)]{Kittel-Q}
Kittel, C., 1963,
{Quantum Theory of Solids}
(New York: Wiley).

\bibitem[Lai(2001)]{Lai01}
Lai, D.
2001,
Rev.\ Mod.\ Phys., 73, 629

\bibitem[Lai \& Salpeter(1997)]{LaiSalpeter97}
Lai, D., \& Salpeter, E.~E.
1997,
\apj, 491, 270

\bibitem[Landau(1930)]{Landau30}
Landau, L.~D.
 1930,
Z.\ f.\ Physik, 64, 629

\bibitem[Landau \& Lifshitz(1977)]{LaLi-QM}
Landau, L.~D., \& Lifshitz, E.~M.
 1977,
{Quantum Mechanics. Non-Relativistic Theory}, 3rd ed.
(Oxford: Pergamon)

\bibitem[Landau \& Lifshitz(1980)]{LaLi-SP1}
Landau, L.~D., \& Lifshitz, E.~M.
1980,
{Statistical Physics, Part 1}, 3rd ed.
(Oxford: Butterworth-Heinemann)

\bibitem[Medin(2012)]{Medin12}
Medin, Z.
2012,
private communication

\bibitem[Medin \& Cumming(2010)]{MedinCumming10}
Medin, Z., \& Cumming, A.
2010,
\pre, 81, 036107

\bibitem[Medin \& Lai(2006)]{MedinLai06}
Medin, Z., \& Lai, D.
2006,
\pra, 74, 062508

\bibitem[Medin \& Lai(2007)]{MedinLai07}
Medin, Z., \& Lai, D.
2007,
\mnras, 382, 1833

\bibitem[Militzer \& Graham(2006)]{MilitzerGraham}
Militzer, B., \& Graham, R.~L.
2006,
J.~Phys.\ Chem.\ Sol. 67, 2136

\bibitem[Morbec \& Capelle(2008)]{MorbecCapelle08}
Morbec, J.~M., \& Capelle, K.
2008,
\prb, 78, 085107

\bibitem[Nagai \& Fukuyama(1982)]{NagaiFukuyama}
Nagai, T., \& Fukuyama, H.
1982,
J.\ Phys.\ Soc.\ Japan, 51, 3431

\bibitem[Nagai \& Fukuyama(1983)]{NagaiFukuyama2}
Nagai, T., \& Fukuyama, H.
1983,
J.\ Phys.\ Soc.\ Japan, 52, 44

\bibitem[Ogata et al.(1993)]{Ogata-ea93}
Ogata, S., Iyetomi, H., Ichimaru, S., \& Van Horn, H.~M.
1993,
\pre, 48, 1344

\bibitem[Paxton et al.(2011)]{MESA}
Paxton, B., Bildsten, L., Dotter, A., et al.
2011,
\apjs, 192, 3

\bibitem[Pollock \& Hansen(1973)]{PollockHansen}
Pollock, E.~L., \& Hansen, J.~P.
1973,
\pra, 8, 3110

\bibitem[Potekhin \& Chabrier(2000)]{PC00}
Potekhin, A.~Y., \& Chabrier, G.
2000,
\pre, {62}, 8554
(Paper~I)

\bibitem[Potekhin \& Chabrier(2004)]{PC04}
Potekhin, A.~Y., \& Chabrier, G.
2004,
\apj, 600, 317

\bibitem[Potekhin \& Chabrier(2010)]{PC10}
Potekhin, A.~Y., \& Chabrier, G.
2010,
Contrib.\ Plasma Phys., 50, 82 (Paper~II)

\bibitem[Potekhin et al.(1999)Potekhin, Chabrier, \& Shibanov]{PCS}
Potekhin, A.~Y., Chabrier, G., \& Shibanov, Yu.~A.
1999,
\pre, 60, 2193
 
\bibitem[Potekhin et al.(2007)Potekhin, Chabrier, \& Yakovlev]{PCY}
Potekhin, A.~Y., Chabrier, G., \& Yakovlev, D.~G.
2007,
\apss, 308, 353
(electronic version with corrected
misprints: arXiv:astro-ph/0611014v3)
 
\bibitem[Potekhin et al.(2009a)Potekhin, Chabrier, \& Rogers]{PCR09}
Potekhin, A.~Y., Chabrier, G., \& Rogers, F.~J.
2009a,
\pre, {79}, 016411

\bibitem[Potekhin et al.(2009b)]{PCCDR09}
Potekhin, A.~Y., Chabrier, G., Chugunov, A.~I., DeWitt, H.~E.,
\& Rogers, F.~J.
2009b,
\pre, {80}, 047401

\bibitem[Potekhin et al.(2012)]{Potekhin-ea12}
Potekhin, A.~Y., Suleimanov, V.~F., van Adelsberg, M., \&
Werner, K.
2012,
\aap, 546, A121

\bibitem[R\"ognvaldsson et al.(1993)]{Rognvaldsson-ea}
R\"ognvaldsson, \"O.~E., Fushiki, I., Gudmundsson, E.~H.,
Pethick, C.~J., \& Yngvason, J.
1993,
\apj, 416, 276

\bibitem[Rosenfeld(1996)]{Rosenfeld96}
Rosenfeld, Y.
1996,
\pre, 54, 2827

\bibitem[Ruderman(1971)]{Ruderman71}
Ruderman, M.~A.
 1971,
\prl, 27, 1306

\bibitem[R\"uster et al.(2006)]{Ruester}
R\"uster, S.~B., Hempel, M., \& Schaffner-Bielich, J.
2006,
\prc, 73, 035804

\bibitem[Salpeter(1961)]{Salpeter61}
Salpeter, E.~E.
1961
\apj, {134}, 669

\bibitem[Schwarz\-schild(1958)]{Schwarzschild}
Schwarzschild, M.
1958,
Structure and Evolution of the Stars
(Princeton: Princeton Univ. Press)

\bibitem[Schwinger(1988)]{Schwinger88}
Schwinger, J.
1988,
Particles, Sources, and Fields 
(Redwood: Addison-Wesley)

\bibitem[Sharma \& Reddy(2011)]{SharmaReddy11}
Sharma, R., \& Reddy, S.
2011,
\prc, 83, 025803

\bibitem[Steinberg et al.(1998)Steinberg, Ortner, \& Ebeling]{Steinberg}
Steinberg, M., Ortner, J., \& Ebeling, W.
1998,
\pre, 58, 3806

\bibitem[Steinberg et al.(2000)Steinberg, Ebeling, \& Ortner]{Steinberg2}
Steinberg, M., Ebeling, W., \& Ortner, J.
2000,
\pre, 61, 2290

\bibitem[Suh \& Mathews(2001)]{SuhMathews}
Suh, I.-S., \& Mathews, G.~J.
2001,
\apj, 546, 1126

\bibitem[Tanaka et al.(1985)Tanaka, Mitake, \& Ichimaru]{Tanaka85}
Tanaka, S., Mitake, S., \& Ichimaru, S.
\pra, {32}, 1896

\bibitem[Thorolfsson et al.(1998)]{Thorolfsson-ea}
Thorolfsson, A., R\"ognvaldsson, \"O.~E., Yngvason, J., \& Gudmundsson, E.~H.
1998,
\apj, 502, 847

\bibitem[Timmes \& Arnett(1999)]{TimmesArnett}
Timmes, F.~X., \& Arnett, D.
1999,
\apjs, 125, 277

\bibitem[Timmes \& Swesty(2000)]{TimmesSwesty}
Timmes, F.~X., \& Swesty, F.~D.
2000,
\apjs, 126, 501

\bibitem[Usov et al.(1980)Usov, Grebenshchikov, \& Ulinich]{UsovGU}
Usov, N.~A., Grebenshchikov, Yu.~B., \& Ulinich, F.~R.
1980,
Sov.\ Phys.--JETP, 51, 148 


\bibitem[van Leeuwen(1921)]{vanLeeuwen}
van Leeuwen, J.~H.
1921,
J.\ de Physique et le Radium, Ser.\ VI, 2, 361

\bibitem[Wickramasinghe \& Ferrario(2000)]{WickramaFerrario00}
Wickramasinghe, D.~T., \& Ferrario, L.
2000,
\pasp, 112, 873

\bibitem[Wigner(1932)]{Wigner32}
Wigner, E.~P.
1932,
Phys.\ Rev. 40, 749

\bibitem[Yakovlev \& Kaminker(1994)]{YakovlevKaminker}
Yakovlev, D.~G., \& Kaminker, A.~D.
1994,
in The Equation of State in Astrophysics,
Proceedings of IAU Colloquium No.~147,
ed.\ G.~Chabrier \& E.~Schatzman
(Cambridge: Cambridge University Press), 214

\bibitem[Yakovlev \& Pethick(2004)]{YakovlevPethick}
Yakovlev, D.~G., \& Pethick, C.~J.
2004,
\araa, 42, 169

\bibitem[Yakovlev \& Shalybkov(1989)]{YaSha}
Yakovlev, D.~G. \& Shalybkov, D.~A.
1989,
Astrophys.\ Space Phys.\ Rev. \textbf{7}, 311

\end{thebibliography}
\end{document}